\documentclass[12pt, onecolumn, draftclsnofoot]{IEEEtran}
\usepackage{amsmath,amsfonts}
\usepackage{algorithmic}
\usepackage{algorithm}
\usepackage{array}
\usepackage[caption=false,font=normalsize,labelfont=sf,textfont=sf]{subfig}
\usepackage{textcomp}
\usepackage{stfloats}
\usepackage{url}
\usepackage{verbatim}
\usepackage{graphicx}
\usepackage{cite}
\usepackage{xcolor}
\usepackage{tikz}
\usepackage{booktabs}
\usepackage{multirow}
\usepackage[]{threeparttable}
\begin{document}

\title{Text Mining Undergraduate Engineering Programs' Applications: the Role of Gender, Nationality, and Socio-economic Status}

\author{Bo Lin,  Bissan Ghaddar,  Ada Hurst
\thanks{B. Lin is with the Department of Mechanical and Industrial Engineering, University of Toronto, Toronto, ON M5S 3G8 Canada (email: blin@mie.utoronto.ca). This work was done when he was with the University of Waterloo.}
\thanks{B. Ghaddar is with Ivey Business School, Western University, London, ON N6G 0N1 Canada (email: bghaddar@uwaterloo.ca).}
\thanks{A. Hurst is with the Department of Management Sciences, University of Waterloo, Waterloo, ON N2L 3G1 Canada (email: adahurst@uwaterloo.ca).}}

\markboth{}%
{Shell \MakeLowercase{\textit{et al.}}: A Sample Article Using IEEEtran.cls for IEEE Journals}


\maketitle

\begin{abstract}
Women, visible minorities, and other socially disadvantaged groups continue to be underrepresented in STEM education. Understanding students' motivations for pursuing a STEM major, and the roles gender, nationality, parental education attainment, and socio-economic background play in shaping students' motivations can support the design of more effective recruitment efforts towards these groups. In this paper, we propose and develop a novel text mining approach incorporating the Latent Dirichlet Allocation and word embeddings to analyze applicants' motivational factors for choosing an engineering program. We apply the proposed method to a dataset of 43,645 applications to the engineering school of a large Canadian university. We then investigate the relationship between applicants' gender, nationality, and family income and educational attainment, and their stated motivations for applying to their engineering program of choice. We find that interest in technology and the desire to make social impact are the two most powerful motivators for applicants. Additionally, while we find significant motivational differences related to applicants' nationality and family socio-economic status, gender has the strongest and the most robust impact on students' motivations for studying engineering.
\end{abstract}

\section{Introduction}
As the defining outcome of national innovation strategies, Canada and other countries around the world have been encouraging students to choose programs in science, technology, engineering, and mathematics (STEM)\cite{Hango}. While it is critical to attract young talents to study STEM, decades of research have also noted the importance of equity, diversity, and inclusion (EDI) in STEM where the underrepresentation of women, visible minorities, and other socially disadvantaged groups has been widely documented \cite{kowtha2008engineering,Macdonald,ASEE2019}. In Canada, among STEM graduates aged 25 to 34, women accounted for only 23\% of those who graduated from engineering, compared with 66\% of university graduates in non-STEM programs \cite{Hango}. Furthermore, persistent lack of gender and racial/ethnic diversity has been reported in STEM-related fields, including but not limited to engineering \cite{kowtha2008engineering,EngineerCanada}, patented invention \cite{hunt2013women,jensen2018gender}, scientific journal editorship \cite{newhouse2021gatekeepers}, and sports analytics \cite{fernandes2022equity}.  

To enhance EDI in the field of STEM, a critical first step is to identify the reasons behind such underrepresentations. Existing studies have categorized the influencing factors into demand- and supply-side factors, corresponding to the action of recruiters and the behaviors of individuals, respectively. On the demand side, negative stereotypes, discrimination, and an unsupportive work/study environment are among the factors that discourage the aforementioned groups from entering and staying in STEM \cite{steinpreis1999impact, hewlett2008athena, grossman2014perceived}. The negative impact of these factors may be amplified by the biases in the algorithms that are used to distribute STEM-related opportunities \cite{lambrecht2019algorithmic}. On the supply side, the lack of role models, the lack of confidence in background knowledge, low perceptions about the social impact of STEM careers, and different life priorities have been identified as key factors that drive the aforementioned groups away from STEM \cite{Besterfield, Marra, hoisl2017sa, samek2019gender}. While these studies have generated insightful findings, they are limited by relatively small sample sizes, which typically originate from surveys and longitudinal studies. To address the challenge, recent studies have leveraged data mining techniques to analyze large datasets originating from more diversed sources, including social media, online job advertisements and university applications, to investigate the gender differences in STEM education \cite{Chopra2} and the challenges confronted by women in STEM occupations \cite{jacobs2020reddit}. Data science technologies present a valuable tool in understanding and promoting EDI in STEM.

In this paper, we present a data-driven analysis of the supply-side factors that motivate students to pursue engineering as a field of study and the roles that various socio-economic factors play in shaping these motivations. Our findings can be used to support future endeavors in promoting EDI in STEM. Specifically, we develop a novel text mining approach to analyze university application data from 43,566 applicants over a period of 5 years. Students' motivations for studying engineering are extracted from their text responses to a series of questions they were required to answer in the application. Motivated by previous findings that students with different demographic and socio-economic backgrounds may be influenced by different factors when making educational choices \cite{Farmer, Wang, Master,bowles2002inheritance,gould2011does,almaas2016willingness}, we then cross-reference the application data with Canadian census data to investigate the impact of gender, nationality, family income, and family educational attainment on shaping these motivations. Our main contributions are summarized below. 
\begin{enumerate}
    \item We propose a novel text mining approach that combines the best of two existing algorithms, Latent Dirichlet Allocation (interpretability) \cite{blei2003latent} and word embedding \cite{Mikolov} (the ability to capture semantic meanings of natural languages). Our approach requires fewer human interventions, compared to existing text mining approaches for analyzing gender differences in STEM, thus is easier to implement and involves less subjectivity. 
    \item Our study is among the first to use text mining approaches to understand the interaction of gender, nationality, family income, and family educational attainment in shaping high-school students' motivation for studying engineering. It adds to the existing literature by providing empirical evidence, originating from a large-scale dataset, that links with education and economic theories.
    \item Our empirical findings yield a rich set of implications for policy making and can be used to support marketing, outreach, and recruitment efforts by university engineering programs, especially for those targeted at underrepresented groups.
\end{enumerate}

\section{Literature Review} \label{sec:lit_review}

\subsection{Motivations for Studying STEM} \label{subsec:motivations}
Prior studies have investigated the influencing factors of students' educational decisions and perceptions, mostly from the social science and education literature. For instance, Beier and Rittmayer \cite{Beier} review the literature on using expectancy-value models to explain STEM educational choices. The model assumes one's educational decisions are shaped by 1) interest, which determines the valence components of expectancy-value models, and 2) self-concept, which is the determinant of both interest and expectation of success for a task. The expectency-value theory can also be used to explain students' adoption of information technology \cite{vanderschaaf2021factors}, and the perception of online learning \cite{lee2021learning}. Alpay et al. \cite{Alpay} explore students' motivations through a cross-faculty survey of undergraduate engineering students at Imperial College London. They find common influences on a student's choice to study engineering include role models, decent salary, and the aspirations to ``invent something new'' and ``make a difference to the world''. These studies provide a list of common factors that impact students' education and career choices, and more importantly, useful frameworks to interpret the text mining results obtained in this study.

Researchers have also investigated the roles that gender and other socio-economic factors play in one's educational choices. Farmer \cite{Farmer} finds that the strength of motivations for women is similar to that of men; however, patterns and types of factors influencing men and women differ significantly. Parent and teacher support, socio-cultural stereotypes, and perceptions about engineers' societal impact are shown to have stronger influences on women than on men \cite{Farmer, Wang, Master}. Bowles and Gintis \cite{bowles2002inheritance}, Gould and Simhon \cite{gould2011does}, and Almas et al. \cite{almaas2016willingness} investigate the inter-generational transmission of economic status and educational attainment. They find the correlation between parental and children's education levels is economically significant. This correlation might be amplified by longer parental exposure and has a stronger effect on boys than on girls \cite{kalil2016father}. Overall, the findings in this research suggest that gender, ethnicity,  and family background affect children's education and socio-economic status in a substantive way, which motivates the analysis in Section \ref{sec:comp_results}.

\subsection{Topic Modeling in Natural Language Procesing} \label{subsec:topic_modelling}

The task of extracting motivations from the university application data is related to topic modeling in natural language processing, which aims to automatically detect word patterns within the given documents that reflect the underlying topics. Classical approaches, including latent semantic indexing \cite{deerwester1990indexing}, probabilistic latent semantic indexing \cite{hofmann1999probabilistic}, and LDA \cite{blei2003latent}, regard each document as a random mixture of some latent topics and each topic as a mixture of words. Once properly trained, one can interpret each topic based on its learned word distribution. Despite their successful applications in various fields, including scientific literature mining \cite{chen2019identify} and patent analysis \cite{wang2019measuring}, these classical approaches are observed to experience performance degradation when applied to short documents due to the lack of document-level word co-occurrence \cite{yi2009comparative,weng2010twitterrank,hong2010empirical}. In our dataset, each applicant's text response consists of, on average, $257$ words, which fall into the category of short documents. Therefore, classical approaches may not be directly applicable.

To tackle the challenge, one promising idea is to integrate word embeddings \cite{Mikolov} with the classical topic modeling approaches. In doing so, topics can be represented and assigned to documents based on the semantic meaning of the texts captured by the word-embedding component, rather than relying on word co-occurrence. One stream of literature focuses on using pre-trained word embeddings to convert discrete words into continuous vector representations and characterizing each topic as a continuous distribution in the embedded semantic space \cite{das2015gaussian, batmanghelich2016nonparametric}, while another stream of literature jointly learns topic and word embeddings in the same semantic space \cite{xu2018distilled, dieng2020topic}. Our approach is similar to the former stream. However, we do not modify the word and topic representations in the LDA training process. Instead, we first fit an LDA model on the given documents and then represent the extracted topics with a pre-trained Word2Vec model \cite{Mikolov} in the semantic space, where we assign topics to each documents based on some distance metrics. This approach is easy to implement and yields results with great interpretability according to our analysis. 

In the context of using text mining to analyze university application data, our approach is related to the method proposed by Chopra et al. \cite{Chopra2}. On a similar dataset, they first utilize a question-answering algorithm to extract sentences from students' text responses that are related to their motivation for studying engineering. The extracted sentences are then mapped to a semantic space and clustered into 200 distinct groups using $k$-Means. They manually merge similar clusters to produce 10 final clusters representing 10 motivational factors. We differ from their approach by replacing the QA extraction phase with an LDA model that extracts common motivations (topics) directly from the application data. As a result, no manual work is needed in merging clusters, which reduces the subjectivity in our analysis.

\section{Method} \label{sec:topic_modeling}
In this section, we first illustrate our motivations for using a topic modeling approach in Section \ref{subsec:prob}. We then present our methodology in Section \ref{subsec:topic_model}.

\subsection{Motivating a topic modeling approach} \label{subsec:prob}
In our dataset, the applicants describe their motivations for applying to their program of choice through a brief text response. Our goal is to 1) extract common motivations that are mentioned by the applicants, and 2) assign each applicant one or several motivations that are mentioned in his/her response. We refer to the first and second items as \textit{motivation extraction} and \textit{motivation assignment} in this paper.

This task has at least three challenges. First, the dataset is large (43,566 applicants), making a manual inspection of all the text responses infeasible. Second, the text data are highly unstructured. Applicants do not explicitly list the reasons/motivations for applying to the engineering school. Instead, the responses are typically organized as a short essay where the author may describe previous experiences, a person they admire, and or career aspirations. As such, algorithms (e.g. those based on keyword counting) that can not capture the semantic meaning of the texts are likely to fail. For example, consider an essay where the applicant introduces a high-school project on robotics design and another essay where the applicant describe her internship at a local hospital. While appearing very different in terms of the vocabularies they use, these two essays could reasonably belong to the same category related to ``previous experience". Third, essays vary in the number of motivational factors included. Therefore, document-level clustering methods that assign only one motivational factor to each text response may under-interpret students' motivations.

To tackle the challenges, we develop a novel text mining approach to perform automatic topic extraction and topic assignment. This approach integrates the LDA and the Word2Vec model to capture the semantic meaning of the text responses and performs detailed sentence-level analysis to more comprehensively capture the motivational factors mentioned by each application. Next, we introduce this approach in detail. 

\subsection{Topic Modeling Approach} \label{subsec:topic_model}

\begin{figure}[!ht]
  \centering
  \includegraphics[width=4in]{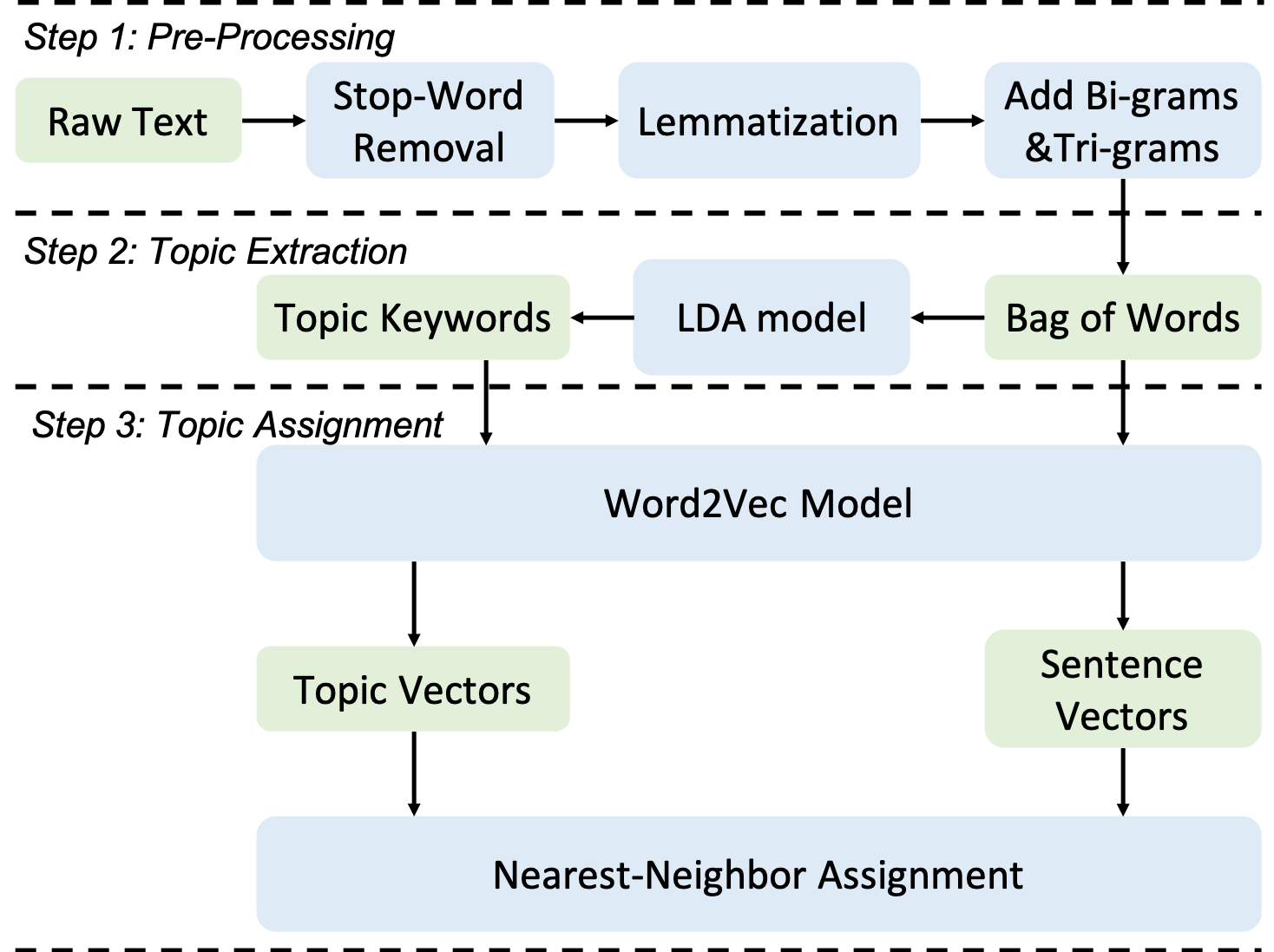}
  \caption{A roadmap of the topic modeling approach.}
  \label{fig:ModelArchitecture}
\end{figure}

We consider the motivation extraction as a topic modeling problem in natural language processing by treating each motivation as a topic to be extracted from the given documents (essays). As illustrated in 
Figure \ref{fig:ModelArchitecture}, this approach consists of three steps. 

\subsubsection{Text Data Pre-processing} 
We first perform standard pre-processing for the text data, including tokenization, stop word removal, and word lemmatization. We extend the set of stop words provided by the Natural Language Toolkit (NLTK) \cite{Bird} by adding words that are frequently used in the essays, including the names of universities and the word ``engineering''. Moreover, we add bi-grams and tri-grams that appear over $30$ times in the dataset to account for important phrases, such as ``co-op'' (co-operative education). Adding bi-grams and tri-grams also helps to extend the length of the essays. This is beneficial for the LDA model, which generally suffers from performance degradation on short documents. At the end of this step, each essay is represented as a bag of words.

\subsubsection{Topic Extraction}
We then employ the LDA model to extract keywords that describe potential topics from the text data. The LDA model takes as input the essays (represented by bags of words) and generates a set of topics and assigns a topic to each essay. Each topic is characterized by a word distribution. When applied to the entire dataset, the extracted topics correspond to the engineering programs because they provide higher variations in the vocabularies used than different motivations. We thus train one LDA model for essays targeted at each engineering program separately and then merge topics across engineering programs based on human interpretation. The number of topics, denoted by $k$, is a hyper-parameter that affects the interpretability of the topic modeling results. We choose this parameter based on the coherent score proposed by \cite{Roder}. We vary the value of $k$ from 1 to 30 inclusively because, according to previous studies \cite{Chopra}, the number of topics in university application essays is within this range. Once the topics are extracted and properly merged, we represent each topic by the words that appear the most frequently in it. We refer to these words as topic keywords in this paper.

\subsubsection{Topic Assignment}
In this step, we first employ the Word2Vec model pre-trained on the Google news corpus \cite{Mikolov} to map all words that appear in the dataset to a 300-dimensional embedding space, where the similarity between the semantic meanings of two words can be measured by the distance between the corresponding word vectors. We then define a topic vector for each extracted topic as the average of the vectors associated with its topic keywords. A sentence vector is computed for each sentence in the essays as the average of the associated word vectors. We then quantify the likelihood of a sentence involving a topic based on the cosine distance between the corresponding topic and sentence vectors. Under the assumption that each sentence involves at most one topic, we assign the nearest topic to each sentence if the distance is below a certain threshold. Setting a distance threshold is necessary because some sentences are irrelevant to the applicant's motivations for studying engineering, according to our preliminary analysis. We aggregate the topics assigned to the sentences in an applicant's essay as his/her motivational factors for studying engineering.

This topic modeling approach generates as outputs a set of motivations (topics) that are commonly mentioned by the applicants to the studied engineering school and a set of motivations that are mentioned by each applicant. By cross-referencing the demographic data of the applicants, we then perform statistical analysis to investigate the impacts of gender, nationality, family income, and family education background in shaping these motivations. Before we proceed to our computational results, we next introduce the university application data and demographic data in detail.

\section{Data} \label{sec:data}
The university application dataset consists of 43,645 applications to 13 direct-entry engineering programs at a large Canadian university spanning 2013-2017. In the application, students were required to indicate their choice of program and write text responses to a list of questions including the reasons for applying to the university and to the specific engineering discipline. A preliminary manual analysis indicated that applicants typically described their motivations for pursuing engineering in their response to the latter question, thus making this text the focus of our analysis. The application system also collected the gender, nationality, and home postal code of the applicants. We cross reference the university application dataset with the income and education data retrieved from the 2016 Canadian census of population based on the home postal codes of the applicants. Next, we present the pre-processing and descriptive statistics of the applicants' demographic data.

\textbf{Gender}. The male-to-female ratio for the entire dataset was $3.22$, indicating, as expected, that engineering is a more popular choice for male than female applicants. When calculated separately for each of the $13$ programs, the ratio ranged from $0.88$ to $7.14$, suggesting a large variation by gender in how attractive different engineering programs are. 

\textbf{Nationality}. Over $70\%$ of the applicants have a Canadian citizenship or permanent residency, hereon referred to as Canadian applicants. The application form does not collect data about the applicant's ethnicity, but it does include the country of origin (by citizenship). A majority of the international applicants come from Asia (45\%) and the Middle East and North Africa (32.5\%), with the remainder originating in North America (8\%), Central and South America (6\%), Sub-Saharan Africa (5.6\%), and Europe (2.9\%). Given the high concentration of international applicants originating from Asia, the Middle East, and North Africa, we use nationality to make some inferences about international students' cultural backgrounds, in order to explain differences between Canadian and international students.

\begin{figure}[!ht]
    \centering
    \includegraphics[width=3in]{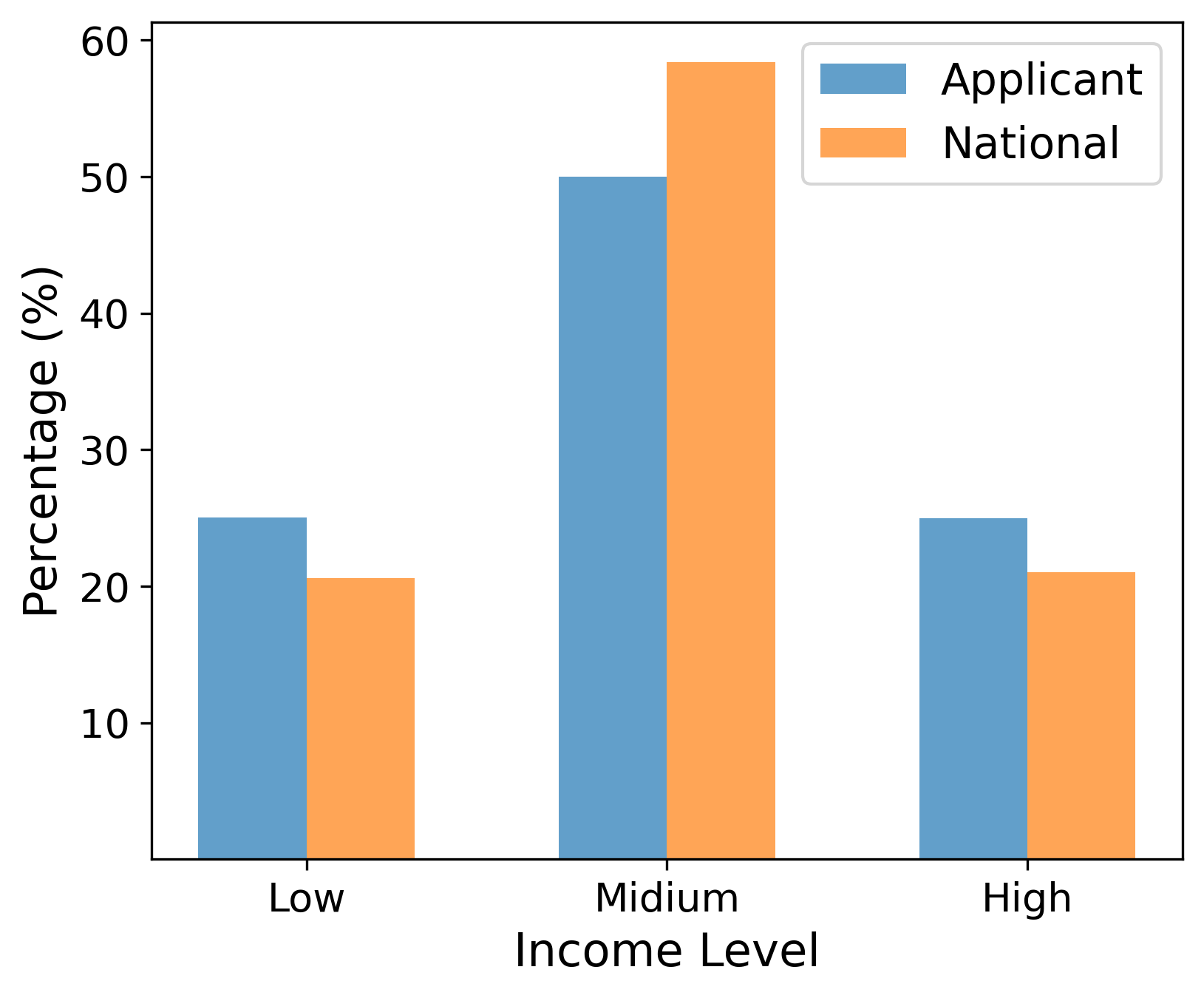}
    \caption{Income distribution for applicant DAs and all Canadian DAs.}
    \label{fig:incomedist}
\end{figure}

\textbf{Family Income}. The 2016 Canadian Census \cite{CanadianCensus} includes average individual before-tax income for dissemination areas (DAs) across Canada. A DA is the smallest standard geographic area for which all census data are disseminated, and it is associated with one or more postal addresses. The home postal codes entered by the applicants enable us to relate them to the corresponding DAs. Assuming that people living in the same community have a similar socio-economic status, we then use the average household income in an applicant's home DA as a proxy for his/her family income. As international applicants typically do not have a Canadian address, we focus solely on Canadian applicants when investigating the impact of family income. The median before-tax income of the applicants' households is 34,992 Canadian Dollars (CAD). 
Applicants are categorized into low-income, mid-income, and high-income groups using the 25\%-percentile value (27,616 CAD/year) and 75\%-percentile value (43,337 CAD/year) of the applicants' family income. 
Figure \ref{fig:incomedist} compares the income distribution of the DAs from which applicants originate to all Canadian DAs, across the three income subgroups. We observe that, compared to the overall Canadian distribution, there are more applicants whose family are in both tails, showing a more dispersed distribution compared to the national average.

\begin{figure}[!ht]
    \centering
    \includegraphics[width=3in]{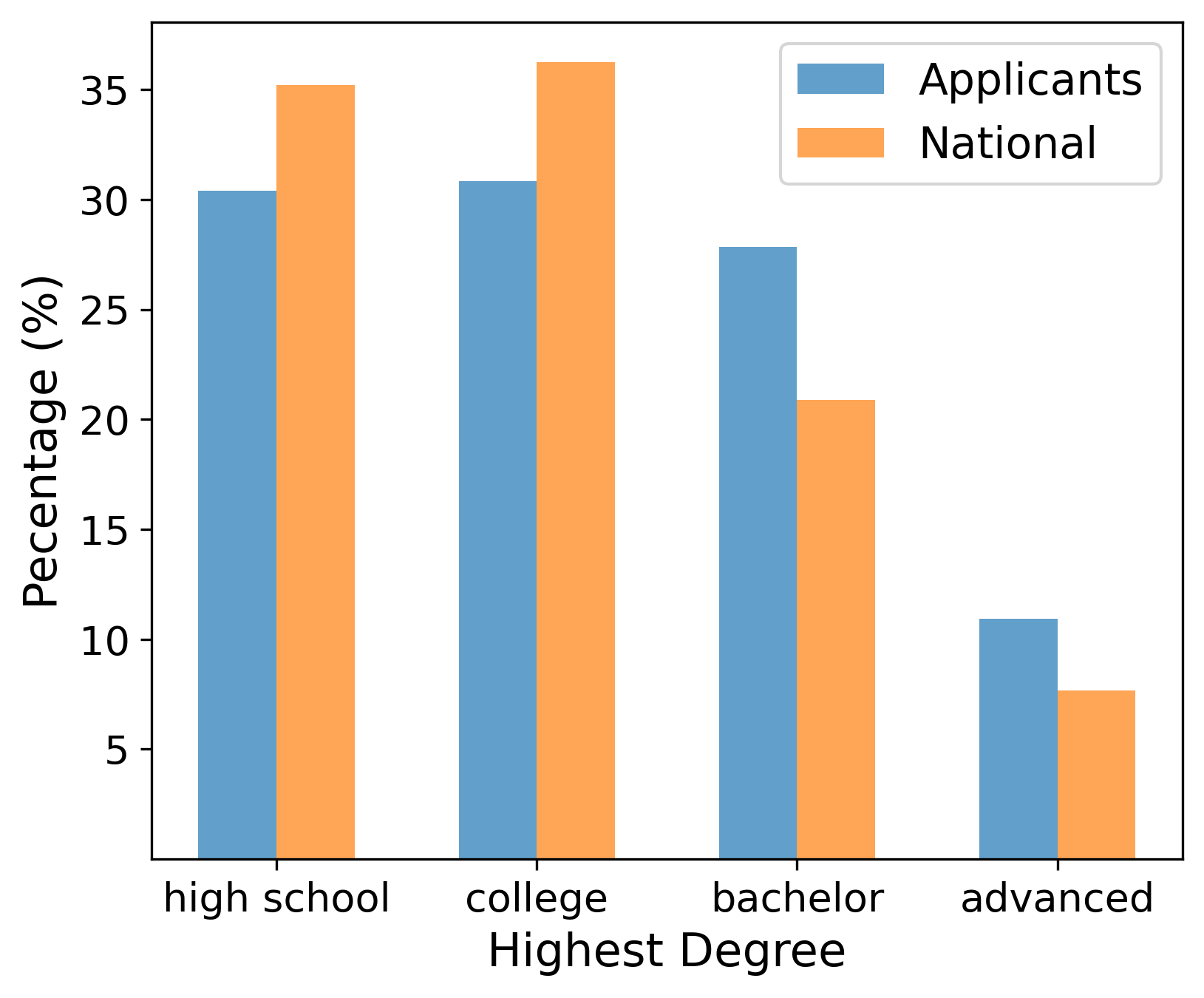} 
    \caption{Highest-degree distributions for applicant DA and all Canadian DAs.}
    \label{fig:HighestDegreeDist}
\end{figure}

\textbf{Family Education}. The 2016 Canadian census provides the number of people aged 25 to 64 years old with different highest certificate, diploma or degree in each DA. For simplicity, we group the education data into four categories: 
1) High school: secondary (high) school diploma, its equivalency certificate or below; 
2) College: certificate and diploma below bachelor level, including those from colleges; 
3) Bachelor: university bachelor degree or any diploma and certificate at bachelor level or above; 
4) Advanced: degree in medicine, dentistry, veterinary medicine, or optometry, master degree and doctorate degree. 
Figure \ref{fig:HighestDegreeDist} compares the highest degree distribution of the DAs from which applicants originate to all Canadian DAs, across the four categories defined above. We observe that compared to the national average, the applicants are more likely to come from DAs where more people have bachelor or advanced degrees. Similar to the analysis done for family income, we group the applicants into two education sub-groups based on the education data associated with their home addresses. Unlike income data, which is uni-dimensional and can thus be categorized based on quantile values, each applicant's family educational background is described by a vector of four entries representing the proportion of people whose highest degree is in the aforementioned four education categories in his/her home DA. We thus employ $k$-Means to cluster the applicants based on their education vectors. The value of $k$ is chosen to be two for interpretability, one for low and one for high education levels. 

\section{Results} \label{sec:comp_results}
In this section, we first present the topic extraction results in Section \ref{subsec:topic_extraction} and then analyze the impacts of gender, nationality, family income, and family education attainment in Sections \ref{subsec:gender}--\ref{subsec:edu}, respectively. Finally, we investigate the interaction of these socioeconomic factors in Section \ref{subsec:interact}.

\subsection{Topic Extraction} \label{subsec:topic_extraction}

\subsubsection{Determining the number of topics} 
\begin{figure}
    \centering
     \includegraphics[width=3in]{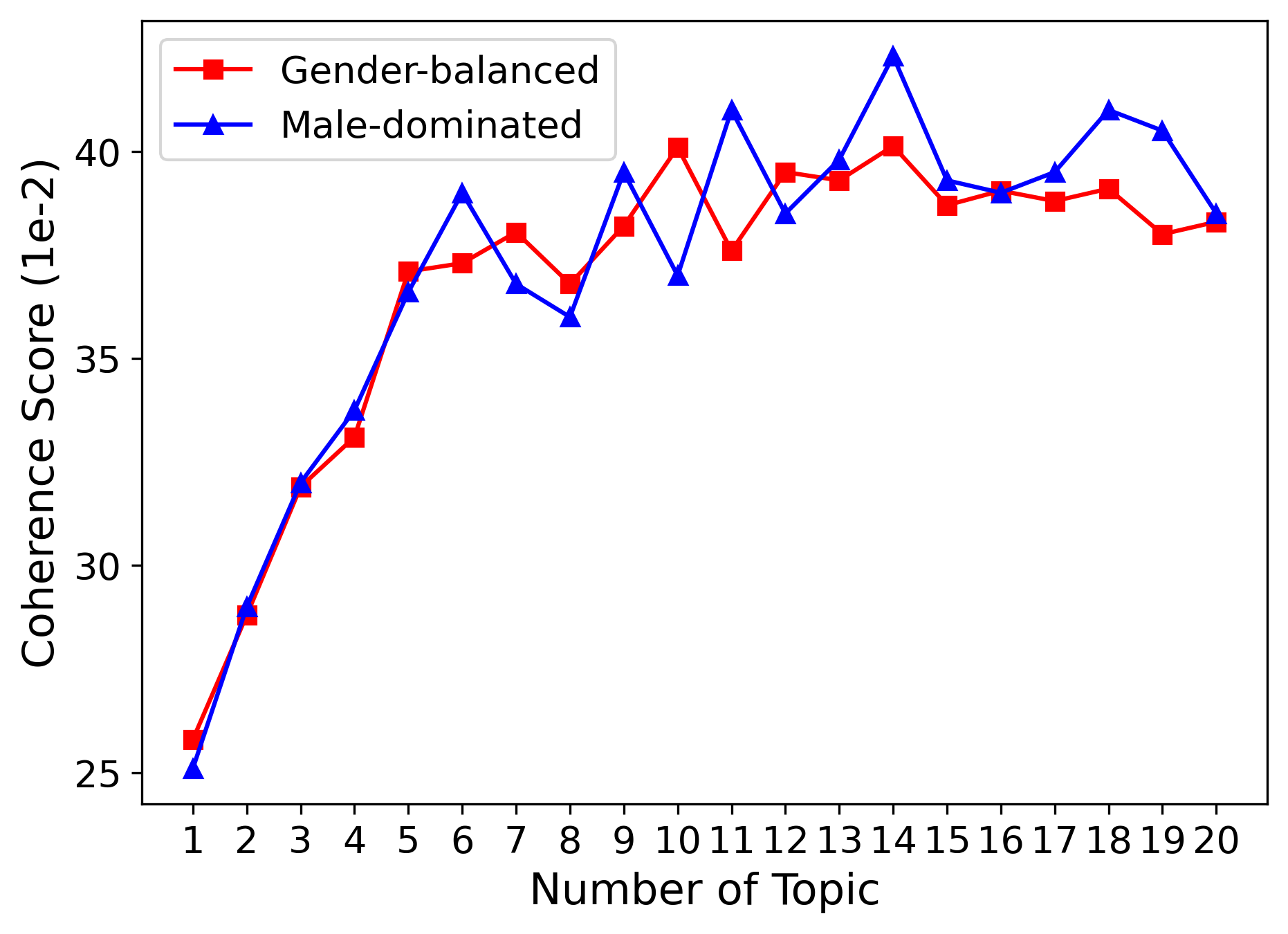}
    \caption{Coherence score by number of topics for a gender-balanced and a male-dominated engineering program.}
    \label{fig:coherencescore}
\end{figure}

We determine the number of topics $k$ based on the coherence scores proposed by \cite{Roder}. A higher coherence score correspond to better interpretability of the topic extraction results. As examples, Figure \ref{fig:coherencescore} illustrates the coherence scores of LDA models trained on essays targeted at a gender-balanced program and a male-dominated program. Both curves have a steep uptrend when $k$ goes from 1 to 7, after which they fluctuate around 0.4. We then manually check the results for $k$ spanning 7 to 12, and choose $k = 10$ based on human interpretation.

\subsubsection{Interpretating the extracted topics} 
Since the topic modeling results for the 13 programs are very similar, we present as an example the extracted topics together with sample texts taken from the gender-balanced program in Table~\ref{tab:LDATopWords}. Essays in all but topics 4, 9 and 10 are closely relevant to the applicants' motivation to study engineering. Applicants in topic 1 aspire to pursue engineering because they enjoy solving real-world problems and would like to learn related skills, while those in topic 2 are attracted to modern technologies such as the development of new medical equipment. In topics 3 and 7, applicants desire to make contributions to society with the help of their knowledge and skills. Similarly, in topic 8, applicants would like to devote themselves to addressing public issues, for instance, environmental pollution. Applicants in topic 5 are interested in math and science and regard engineering as an extension of these subjects. Finally, applicants in topic 6 view engineering study as a cornerstone for the fulfillment of their career goals.

\begin{table}[!ht]

\caption{LDA topics with top words and sample texts}
\centering
\renewcommand{\arraystretch}{0.5}
\footnotesize{
\begin{tabular}{@{}cl@{}}
\toprule
Number                        & \multicolumn{1}{c}{Top Words \& Sample Text}                                                                                                                                                                                                               \\ \midrule
\multirow{2}{*}{1}            & \textbf{Problem, solve, skill, solution, problem solve, design, world, real, math}                                                                                                                                                                         \\ \cmidrule(l){2-2} 
                              & \textit{\begin{tabular}[c]{@{}l@{}}I consider myself an analytical person who takes interest in analyzing and solving problems. \end{tabular}}                                                                                  \\ \midrule
\multirow{2}{*}{2}            & \textbf{Medical, technology, bio, interested, new, help, doctor}                                                                                                                                                                                           \\ \cmidrule(l){2-2} 
                              & \textit{\begin{tabular}[c]{@{}l@{}}I was amazed by the new technologies we were exposed to and it sparked my interest in engineering as \\ it was constantly changing and evolving.\end{tabular}}                                                          \\ \midrule
3                             & \textbf{Life, health, improve, technology, people, help, human, develop, care, world}                                                                                                                                                                      \\\cmidrule(l){2-2}
                              & \textit{\begin{tabular}[c]{@{}l@{}}I really enjoy helping those who are in need and would absolutely love to create equipment that would \\ assist doctors and nurses to help patients.\end{tabular}}                                                       \\ \midrule
4                             & \textbf{Design, build, create, project, use, human, technology, body, machine, robotics}                                                                                                                                                                   \\ \cmidrule(l){2-2} 
                              & \textit{\begin{tabular}[c]{@{}l@{}}My passion grew as I joined engineering club at my school. We had a lot of fun making towers and \\ competing to make the strongest paper bridge as we had to be efficient and make wise decisions.\end{tabular}}       \\ \cmidrule(l){2-2} 
\multicolumn{1}{l}{}          & \textit{\begin{tabular}[c]{@{}l@{}}At the Elite Engineer Mentorship program I worked in a lab that produced biofuels from  waste oil.  \\ My discussions with engineers and students in these programs have inspired me to pursue this field.\end{tabular}} \\ \cmidrule(l){2-2}
                              & \textit{\begin{tabular}[c]{@{}l@{}}I have always had an interest in building things since childhood when my dad used to take me to a \\ kids workshop at the Home Depot, where we built toy models.\end{tabular}} \\\midrule
5                             & \textbf{Physic, biology, math, interested, chemistry, subject, apply, study}                                                                                                                                                                               \\ \cmidrule(l){2-2} 
                              & \textit{\begin{tabular}[c]{@{}l@{}}I am very strong and interested in math and sciences and Biomedical Engineering is a perfect \\ combination of all the subjects that I love.\end{tabular}}                                                              \\ \midrule
6                             & \textbf{Career, goal, skill, help, pursue, passion, believe, allow, others}                                                                                                                                                                                \\ \cmidrule(l){2-2} 
                              & \textit{\begin{tabular}[c]{@{}l@{}}It was clear that engineering can lead me to the careers that are ultimately meaningful and is \\ well paid for.\end{tabular}}                                                                                          \\ \midrule
7                             & \textbf{Want, help, make, people, world, thing, life, love}                                                                                                                                                                                                \\ \cmidrule(l){2-2} 
\multicolumn{1}{l}{\textbf{}} & \textit{\begin{tabular}[c]{@{}l@{}}Engineering is something we can apply to our society, and I want to get that satisfaction when we \\ can see the results by helping others.\end{tabular}}                                                               \\ \midrule
8                             & \textbf{Environment, water, want, issue, pollution, world, problem, interested, make}                                                                                                                                                                      \\ \cmidrule(l){2-2} 
                              & \textit{The rise in environmental issues and a low awareness of them have motivated me to take action.}                                                                                                                                                     \\ \midrule
9                             & \textbf{School, high, high school, grade, course, physic, year, take, career, math}                                                                                                                                                                        \\ \cmidrule(l){2-2} 
                              & \textit{\begin{tabular}[c]{@{}l@{}}When I was in high school, I took the activities about robot making and remote coding which improve\\ my practical skills and enriched the knowledge about technology.\end{tabular}}                        \\ \cmidrule(l){2-2} 
\multicolumn{1}{l}{\textbf{}} & \textit{\begin{tabular}[c]{@{}l@{}}I have had multiple conversations with my teachers at school discussing environmental engineering \\ as a potential career choice and have been greatly encouraged to pursue it.\end{tabular}}                          \\ \midrule
10                            & \textbf{University, student, co-op, opportunity, study}                                                                                                                                                                                                    \\ \cmidrule(l){2-2} 
                              & \textit{\begin{tabular}[c]{@{}l@{}}Combining the academic superiority with the world-renowned cooperative education program, this \\ degree is what will allow me to become the engineer and entrepreneur that I know I can be.\end{tabular}}                                    \\ \bottomrule
\end{tabular}}
  \label{tab:LDATopWords}
\end{table}

\begin{table}[!ht]
\centering
\renewcommand \arraystretch{0.7}
\caption{List of topics with keywords and threshold values}
\footnotesize{
    \begin{tabular}{@{}cllc@{}}
        \toprule
        Number & \multicolumn{1}{c}{Topic} & \multicolumn{1}{c}{Keywords}                                                     & Threshold \\ \midrule
        1      & Mentorship                & Teacher, father, family, uncle, brother                                          & 0.40      \\
        2      & Public Issue Concerns     & Issue, environment, pollution, protect, sustainability, energy, disaster, health & 0.60      \\
        3      & Childhood Dream           & Dream, childhood, young                                                          & 0.50      \\
        4      & Professional Development  & Career, path, goal, skill, opportunity, future                                   & 0.50      \\
        5      & Love of Science           & Physics, math, chemistry, biology, science, interest                             & 0.60      \\
        6      & Previous experience       & Project, competition, experience, club, grade, school, team                      & 0.50      \\
        7      & Societal Contribution    & Improve, people. quality, life, contribution, society, help, others             & 0.50        \\
        8      & Technical Interest        & Develop, design, create, invent, technology, equipment, research                 & 0.55      \\
        9      & Problem Solving           & Solve, problem, apply, knowledge, challenge                                      & 0.60      \\
        \bottomrule
    \end{tabular}}
\label{tab:keywords}
\end{table}

The connections of topics 4, 9 and 10 to the applicants' motivation for studying engineering are harder to interpret. In topic 4, most of the responses relate to the applicants' previous experiences (e.g., successfully designing a paper bridge, taking part in a research project, etc.), which can be easily recognized from its keywords. Some other responses put greater emphasis on the people who offer supervision and guidance through their experience, while others also relate their motivation to childhood experiences, indicating that becoming an engineer is their treasured childhood dream. In this case, we split topic 4 into three as the applicants are not only motivated by their previous experiences, but also by mentorship and childhood dreams, which are considered different motivational factors. In topic 9, the phrase ``high school'' is frequently mentioned. Nevertheless, we do not regard ``high school'' in itself as a source of motivation. It is the experiences or mentorship received during this period of time that actually drives students to pursue the study of engineering. Hence, we do not include ``high school'' in our final topic list. In topic 10, most applicants discuss the university's reputation and some program specific information, which are also irrelevant to their motivation to becoming engineers but rather specific to the university they are applying to. 

Table \ref{tab:keywords} presents the list of $9$ final topics with the corresponding keywords. Most of the keywords come from the top words generated by the LDA, while others are manually selected as an example from the given responses. Due to the similarity of the programs' topic modeling results, in the following analysis, we use the same set of keywords for all programs. As mentioned earlier, to avoid assigning topics to irrelevant sentences, we set a threshold for each topic center through experiments with sample sentences from applicants' responses. A sentence is assigned to a topic only when its calculated cosine distance between the sentence and topic vectors exceeds the corresponding threshold.

\subsubsection{Motivation distribution} 
\begin{figure}[!ht]
    \centering
    \includegraphics[width=3.3in]{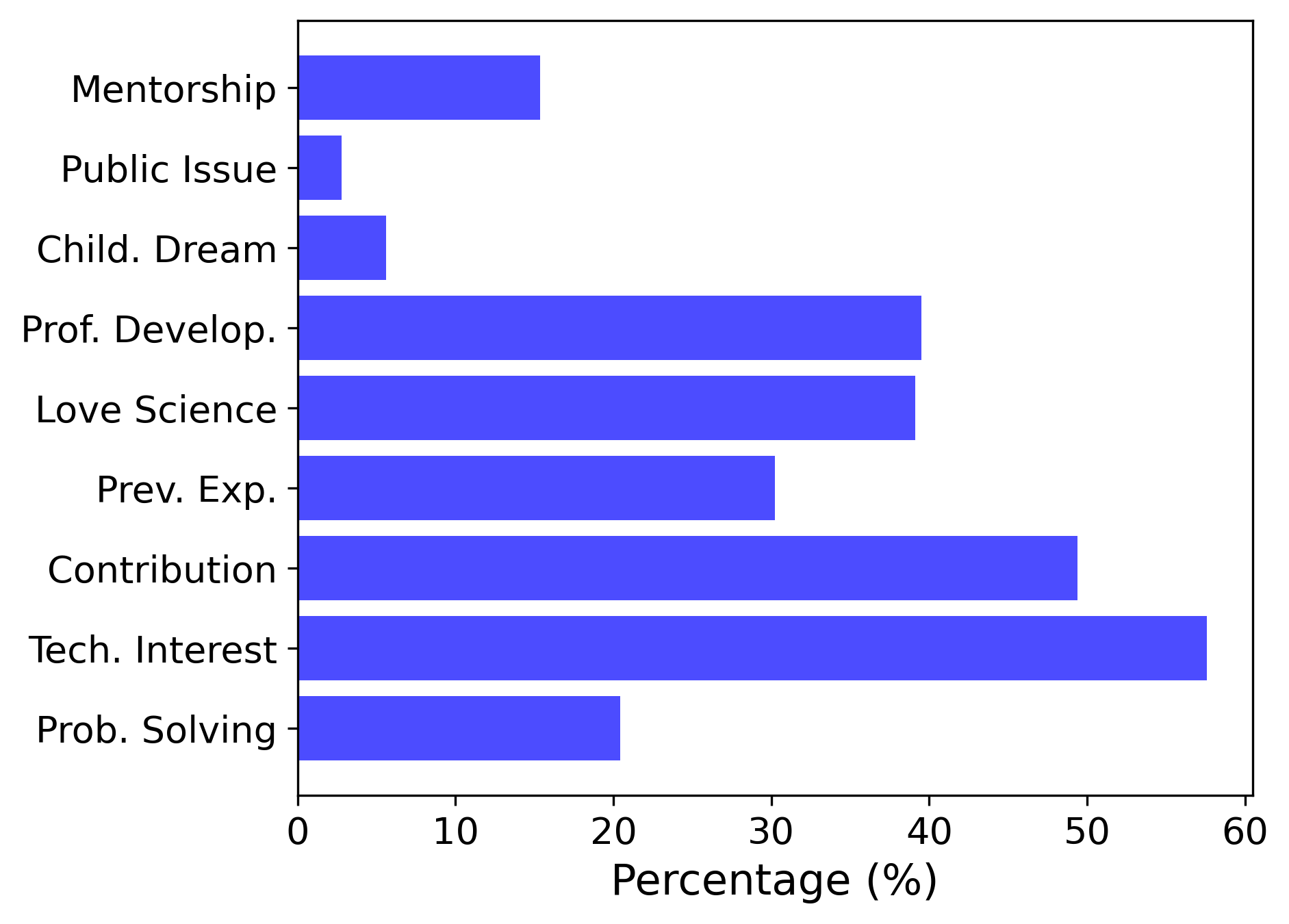}
    \caption{The percentage of applicants who mention each motivational factor.}
    \label{fig:topic_dist}
\end{figure}

Figure \ref{fig:topic_dist} presents the frequency that each motivational factor was mentioned by the applicants. Among the factors, ``Interest in technology'' was the most common, followed by ``making societal contributions'', which was mentioned by nearly 50\% of the applicants. Over 30\% of the applicants stated ``professional development'', ``love of science'', and ``previous experience'' as factors that motivated them to study engineering. Next, we perform statistical analyses to investigate the roles of different demographic factors in shaping these motivations.

\subsection{Impacts of Gender} \label{subsec:gender}

\begin{figure}[!ht]
    \centering
    \includegraphics[width=4in]{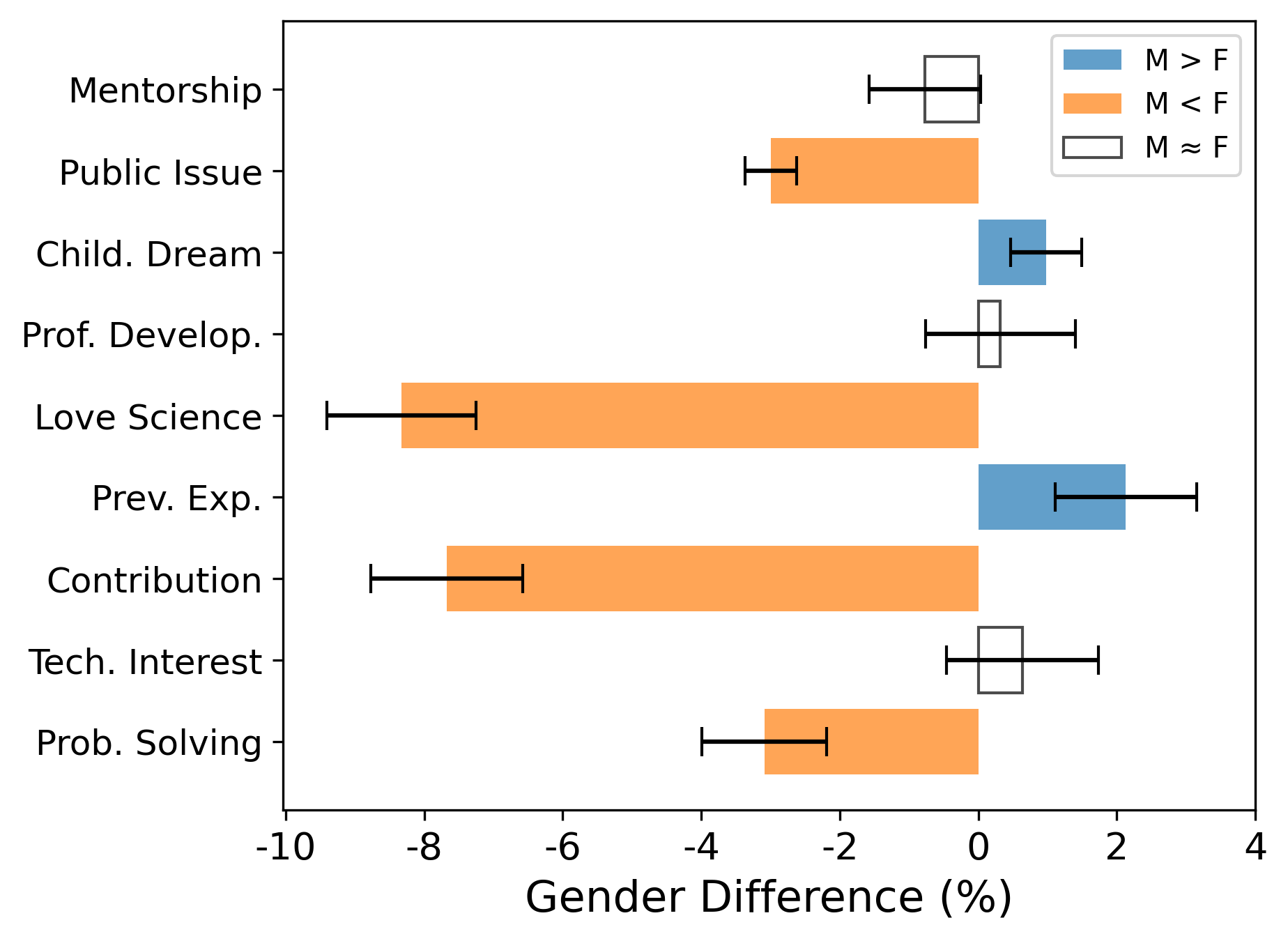} 
    \caption{Gender-based differences in topic frequencies. Female- and male-dominated topics are highlighted in orange and blue, respectively. The error bars indicate the 95\% confidence intervals.}
    \label{fig:gender_all}
\end{figure}

We quantify the gender difference by computing, for each motivational factor, the difference between the percentages of male and female applicants who mentioned the factor. We then perform significance tests on the calculated differences and summarize the results in Figure \ref{fig:gender_all}. 

A significantly larger proportion of males than females described their motivations for applying to engineering as linked to their early childhood aspirations (``childhood dream'') and related experiences with STEM-related activities (``previous experience''). In contrast, female applicants placed greater emphasis on finding solutions to issues of public concern (``public issue'' and ``problem solving'') and thus contributing to societal well-being (``societal contribution''). Female applicants are also significantly more likely to mention their love of math, physics, and life sciences (``love of science''). We do not observe any statistically significant gender differences in the topics related to ``mentorship'', ``technical interest'', and ``professional development''.

\subsection{Impacts of Nationality} \label{subsec:nation}
\begin{figure}[!ht]
    \centering
    \includegraphics[width=4in]{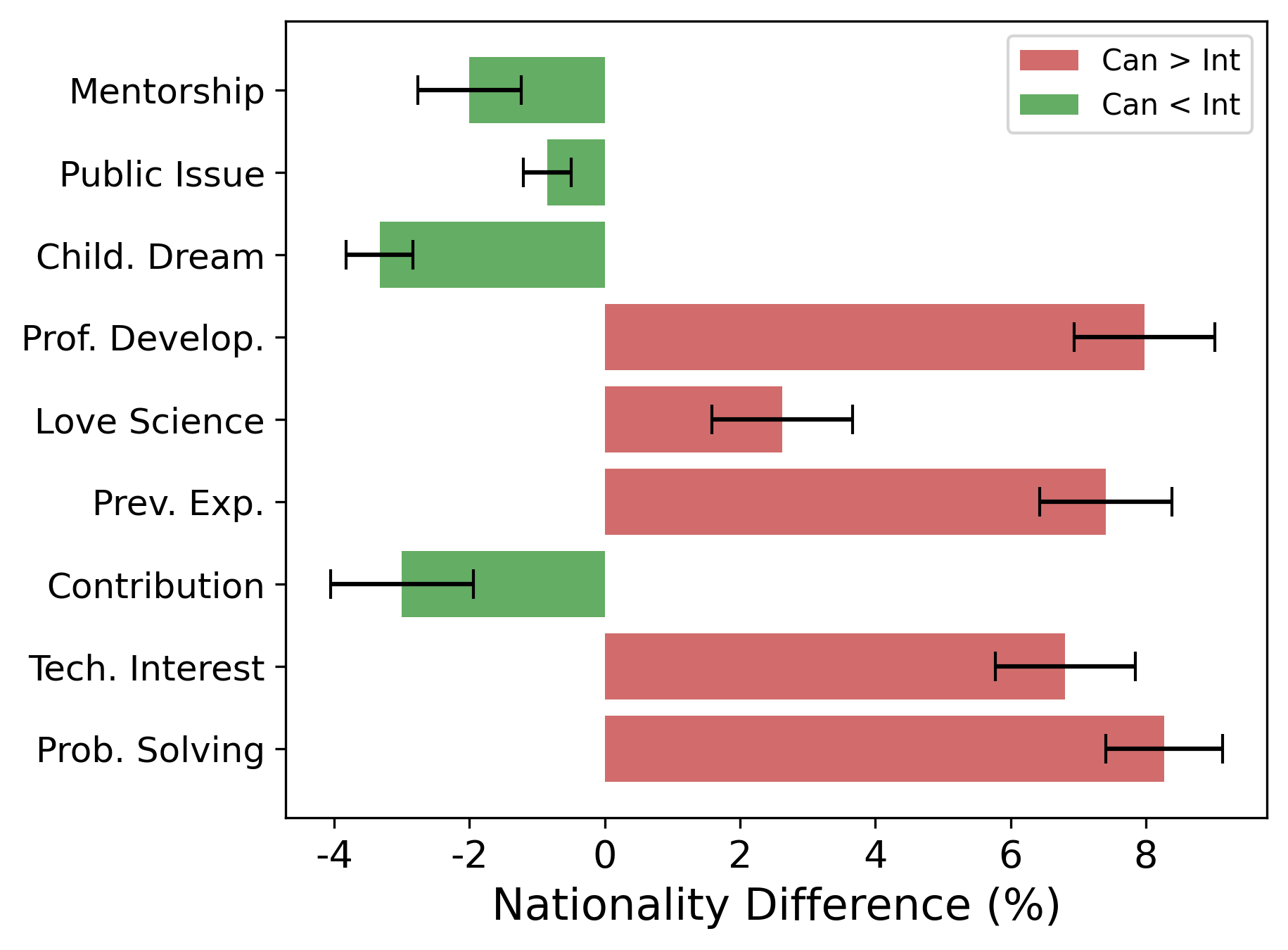} 
    \caption{Nationality-based differences in topic Frequencies. Topics dominated by Canadian and international applicants are highlighted in red and green, respectively. The error bars indicate the 95\% confidence intervals.}
    \label{fig:nationality_all}
\end{figure}

We adopt a similar approach as in the previous section to quantify the nationality difference in each motivational factor as the difference between the percentages of Canadian and international applicants who mentioned the factor. The computational results are summarized in Figure 
\ref{fig:nationality_all}.

For all motivational factors, we find significant differences by nationality at a confidence level of 5\%. Canadian applicants are significantly more likely to mention ``professional development'', ``love of science'', ``previous experience'', ``technical interest'',  and ``problem solving'', while more international applicants have their motivations clustered under the topics of ``mentorship'', ``public issue'', ``childhood dream'', and ``contribution''. A likely interpretation of these differences is that Canadian applicants are more driven by ``individualistic'' factors. They choose to pursue engineering as a field of study due to their interests in technology and science, personal enjoyment of problem solving, previous experiences, and the expected promising career paths for engineering graduates. In contrast, the data suggests that international applicants put greater emphasis on the ``collective good'' (as also discussed in \cite{triandis2018individualism}), driven by concerns around public issues and the opportunity to make societal contributions. The aspirations they establish at young ages as well as the presence of role models who provide mentorship in relevant fields are also more important to them than to Canadian applicants.

\subsection{Impacts of Family Income} \label{subsec:income}
\begin{figure}[!ht]
    \centering
    \includegraphics[width=4in]{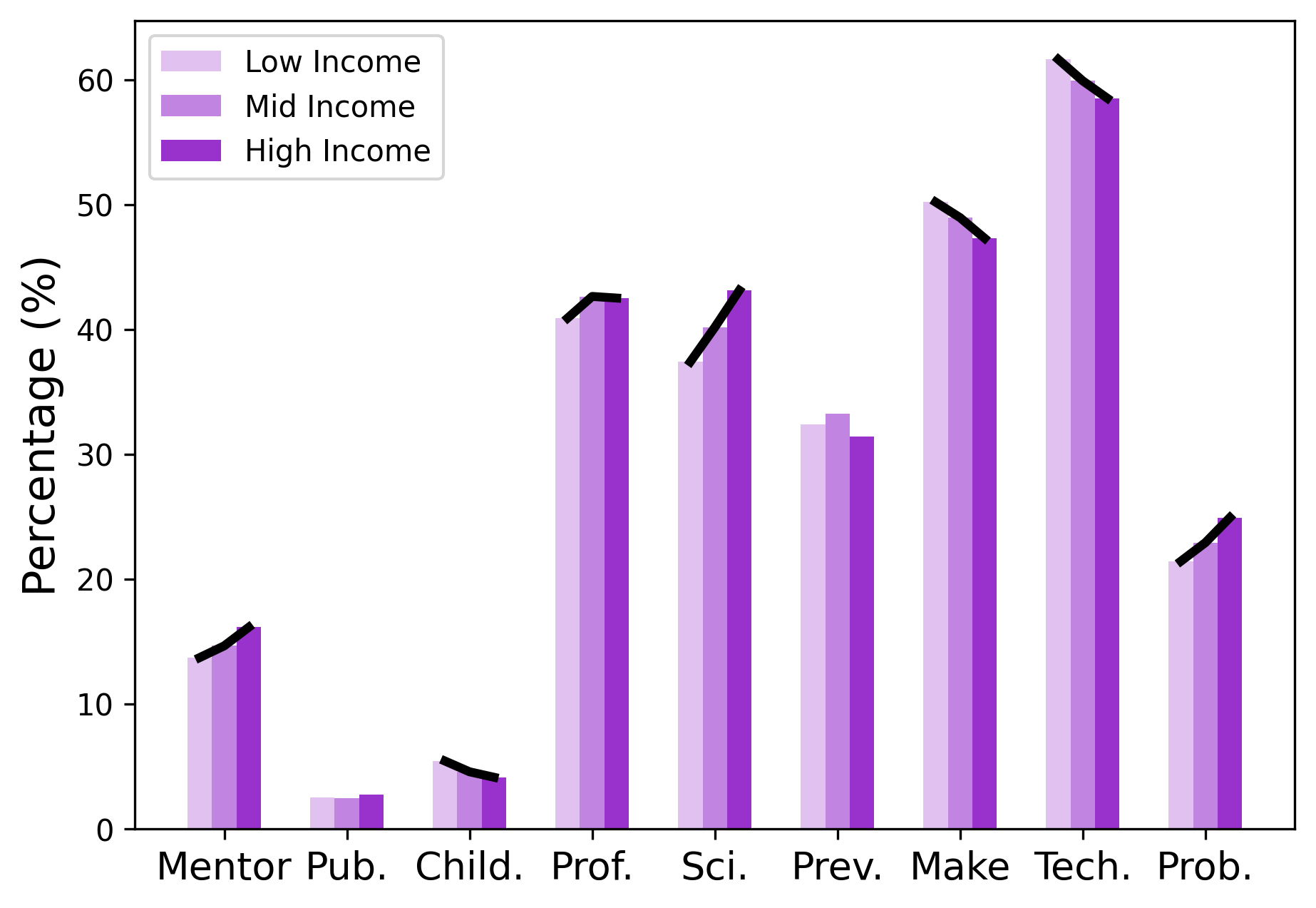} 
    \caption{Family income-based differences in Topic frequencies. Differences that are significant at a confidence level of 95\% are highlighted with trend lines.}
    \label{fig:income_trend}
\end{figure}

Next, we investigate how applicants' family income might affect their motivation for applying to an engineering program. Figure \ref{fig:income_trend} summarizes the frequencies that each topic was mentioned by applicants from the income groups and the difference among them.  

We observe\textit{} that the higher an applicant's household income, the greater the probability that they would mention reasons clustered in the topics of ``problem solving'', ``love of science'', ``professional development'', and ``mentorship''. In contrast, increasing  household income is associated with a decreased frequency in the topics of ``technical interests'', making ``contributions'' to society and ``childhood dreams''. These results are consistent with the findings presented in \cite{valadez1998applying} that students from families with higher socio-economic status usually have access to resources, such as high quality science education, exposure to relevant extracurricular activities, and proper guidance regarding career development, that may help to shape decisions about program selection. Students from low income families on the other hand, are generally less likely to take advantage of the aforementioned resources. Therefore, they may pay more attention to modern technology and the social implications of technology that they can easily perceive through in-class learning and web sources.

\subsection{Impacts of Family Education Attainment} \label{subsec:edu}
\begin{figure}[!ht]
    \centering
    \includegraphics[width=3.5in]{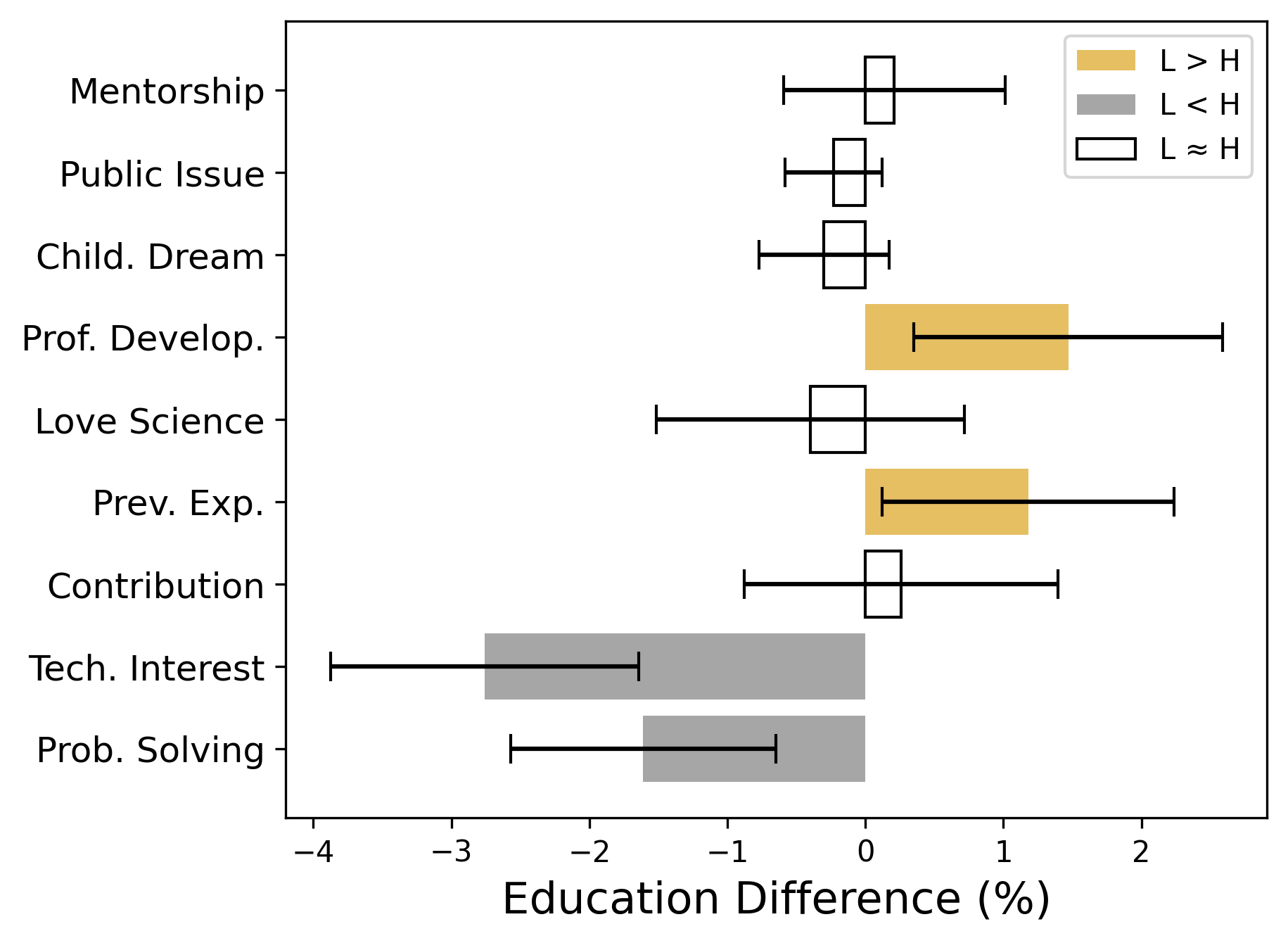} 
    \caption{Education-based differences in topic frequencies. Topics dominated by applicants from high- and low-education families are highlighted in grey and yellow, respectively. Error bars indicate 95\% confidence intervals.}
    \label{fig:education_all}
\end{figure}

We then examine differences in motivational factors due to family education. The results are summarized in Figure \ref{fig:education_all}. We find that overall there are only few education-based significant differences in the frequency of each topic. Applicants whose families have higher education are significantly more likely to mention ``technical interests'' and the enjoyment of ``problem solving'', and less likely to mention the topics of ``previous experience'' and ``professional development''.  

\subsection{Interaction of Gender, Nationality, Family Education Attainment, and Family Income}  \label{subsec:interact}
Next, we investigate the intersections among the demographic factors. We mainly focus on gender verses other factors because family income and education data were available only for Canadian applicants. We also extract insights around the interaction between family income and education backgrounds for Canadian applicants.

\subsubsection{Gender and Nationality}

\begin{figure}[!ht]
    \centering
    \begin{minipage}[t]{0.48\textwidth}
        \includegraphics[width=3in]{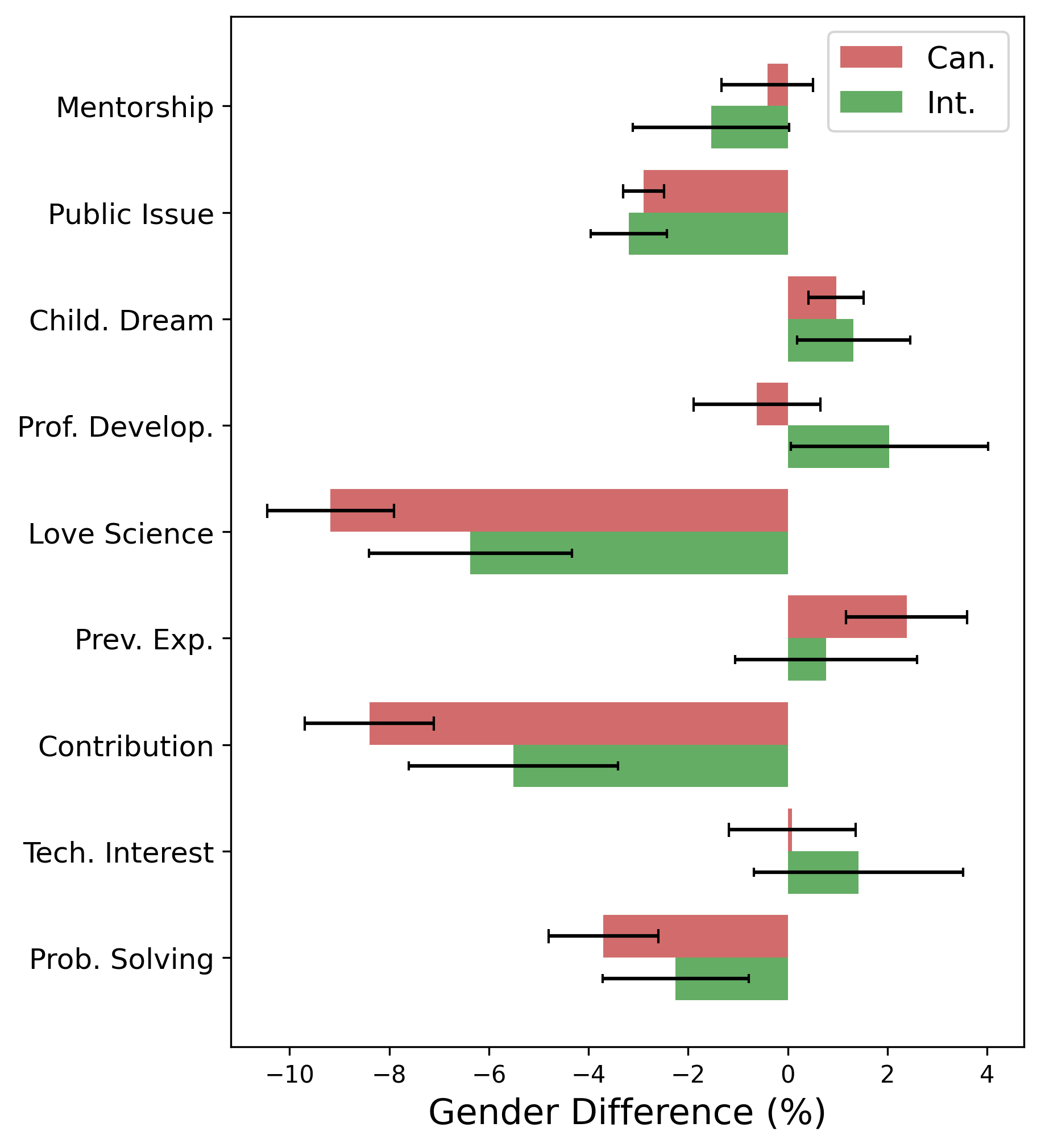} 
        \caption{Gender difference by nationality. Error bars indicate 95\% confidence intervals.}
        \label{fig:Gender_by_Nation}
    \end{minipage}
    \begin{minipage}[t]{0.48\textwidth}
        \includegraphics[width=3in]{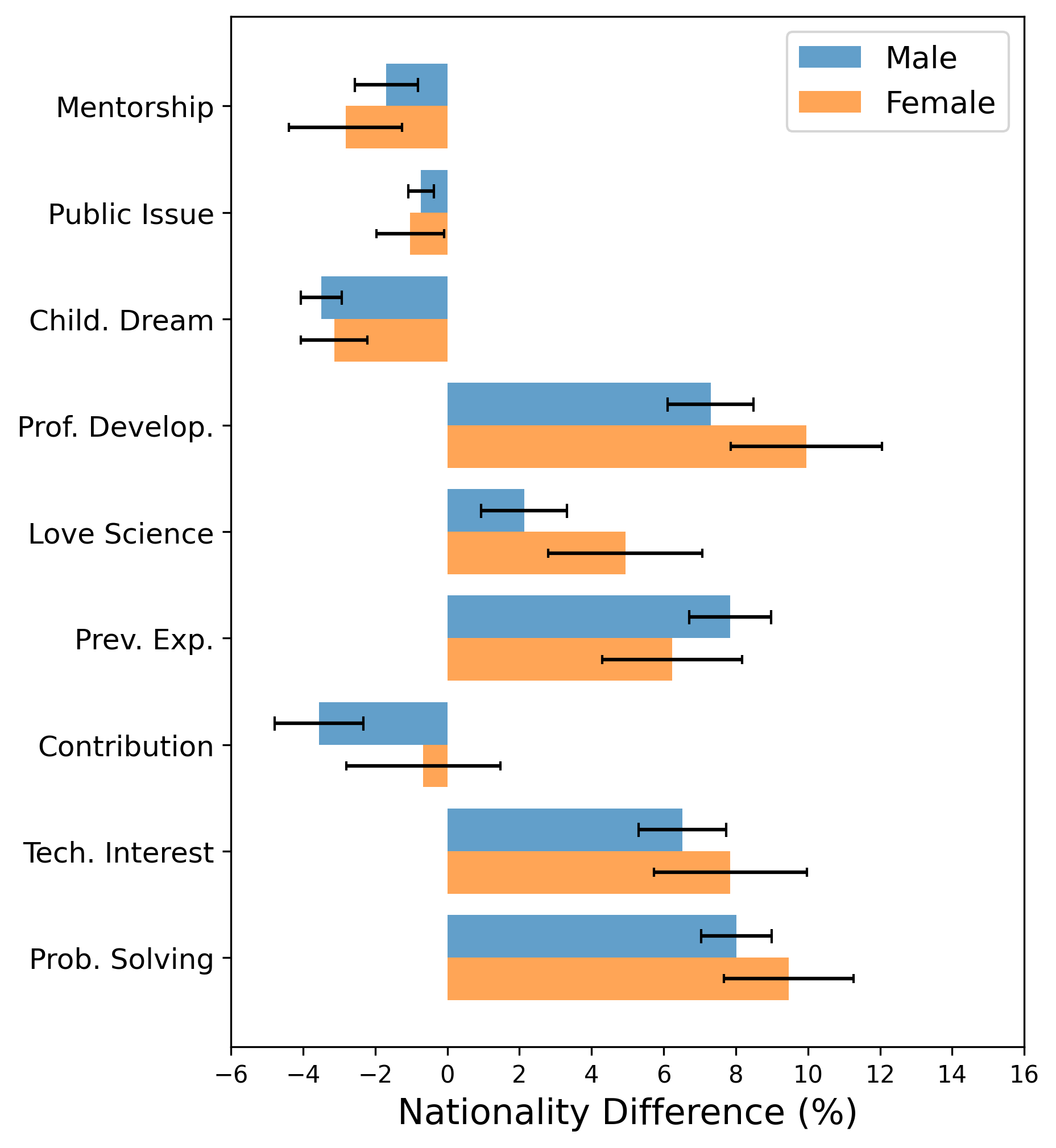} 
        \caption{Nationality difference by gender. Error bars indicate 95\% confidence intervals.}
        \label{fig:Nation_by_Gender}
    \end{minipage}
\end{figure}

We first compare topic frequencies by gender within the Canadian and international applicant sub-groups, as shown in Figure \ref{fig:Gender_by_Nation}. Overall, gender differences are mostly consistent in both Canadian and International applicant subgroups. The only two distinctions between Canadian and international applicants lie in the frequencies of the ``professional development'' and ``previous experience'' topics. We previously reported that there was no significant gender-based difference in the frequency of the ``professional development'' topic ($0.32\%$); however, when looking at international applicants only, males mention this topic more than their female counterparts (difference of $2.04\%$). The difference is insignificant and has the opposite direction for Canadian applicants. In addition, we also previously found that male applicants are more likely to mention the topic of ``previous experience'' than females (difference of $2.13\%$); however it appears that the difference is driven primarily by Canadian applicants (for whom the difference between genders is $2.39\%$). No significant gender-based difference is found in international applicants for this topic. 

We also compare topic frequencies by nationality within the female and male applicant subgroups, as shown in Figure \ref{fig:Nation_by_Gender}. We observe that overall, nationality-based differences persist regardless of gender. The only exception appears to be the topic of ``make contribution''. While we previously found that international students were more likely to mention this topic than Canadian ones (difference of $3.00\%$), the difference appears to be driven by male applicants (difference of -$3.56\%$); no significant nationality-based difference is found within female applicants.

All together, this analysis indicates that with very few exceptions, there is minimal interaction between gender and nationality in applicants' motivations for applying to engineering programs.

\subsubsection{Gender and Family Income}

\begin{figure}[!ht]
\centering
\includegraphics[width=3.3in]{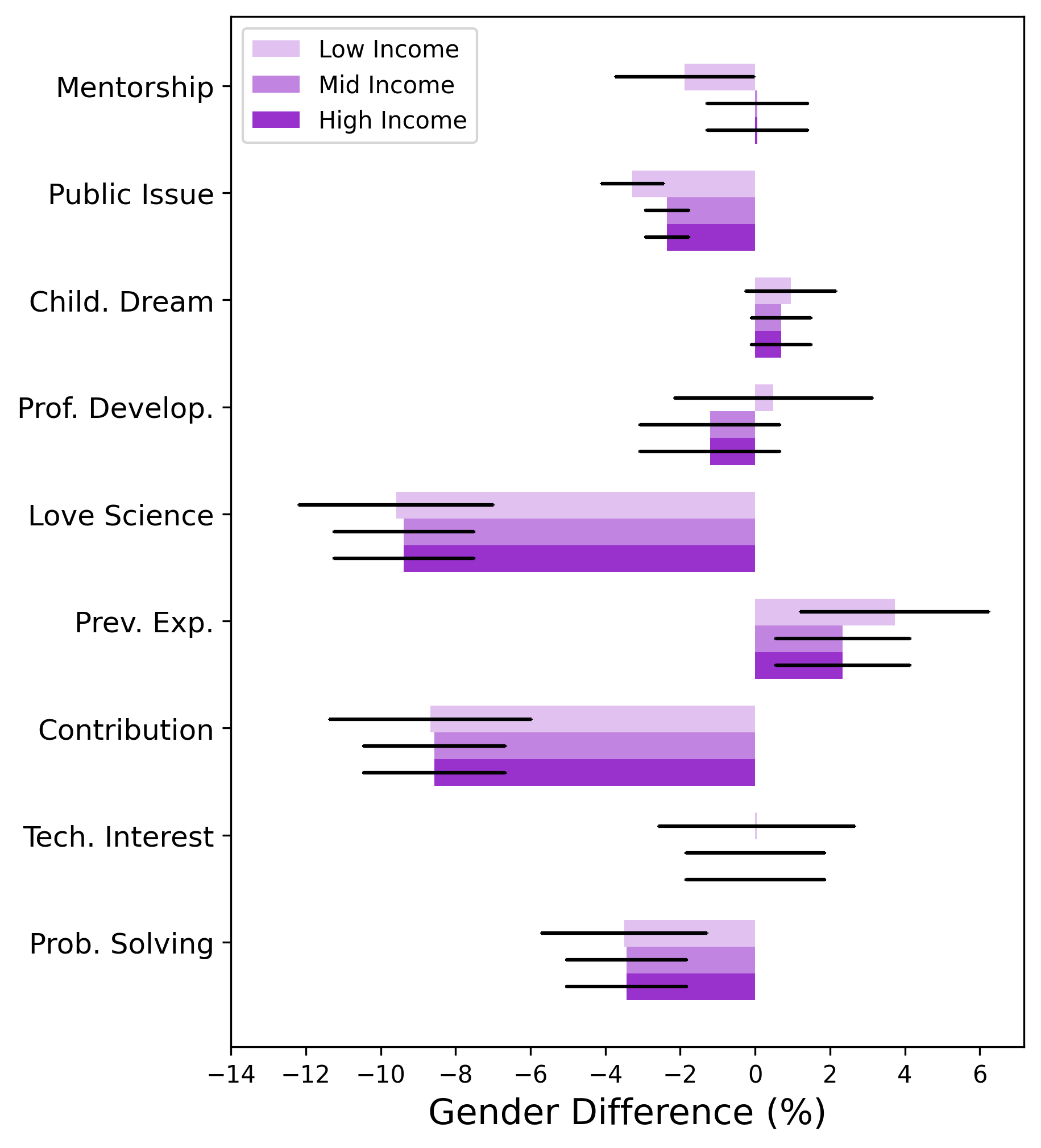} 
\caption{Gender difference by family income. Error bars indicate 95\% confidence intervals.}
\label{fig:Gender_by_Income}
\end{figure}

We first compare topic frequencies by gender within the low, mid, and high income subgroups, as shown in Figure \ref{fig:Gender_by_Income}. While slight differences are noted between income groups, overall, gender differences in topics are very robust across income levels. The only notable difference lies in the topic of ``mentorship'', for which a significant gender difference (-1.88\% - suggesting it's more frequent with females) is only found in the low-income group of applicants. 

\begin{figure}[!ht]
\centering
\includegraphics[width=6.5in]{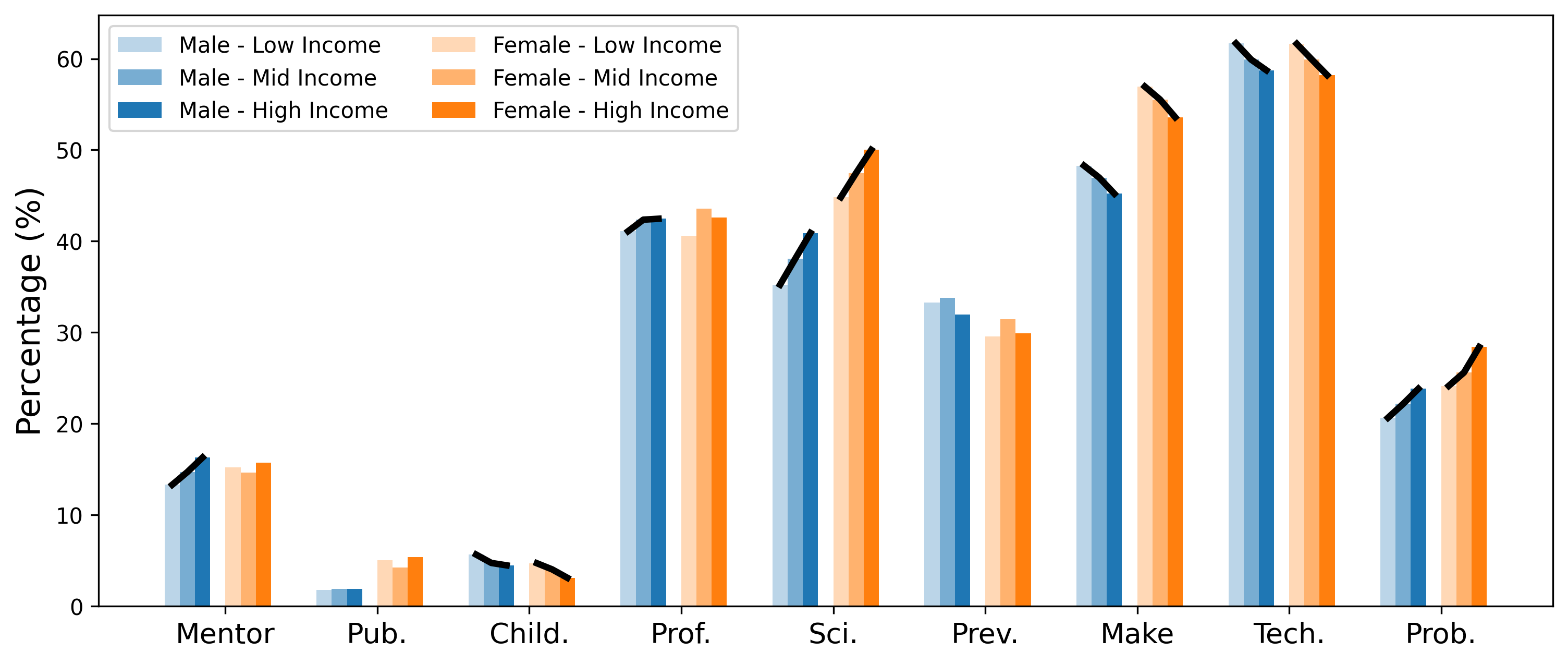} 
\caption{Topic distribution by gender and income. Income differences that are significant at a confidence level of 5\% are highlighted with trend lines.}
\label{fig:Income_by_Gender}
\end{figure}

We next compare topic frequencies by income separately for male and female applicants (Figure \ref{fig:Income_by_Gender}). The patterns of ``problem solving'' and ``love of science'' subject to income levels remain clear regardless of gender, while the significance of other topics are diminished to some extent indicating that the income difference in these topics can be partially explained by the gender distribution across the income levels. This is also consistent with the findings in \cite{almaas2016willingness, bertrand2013trouble} as the results imply that family income has a greater influence on males than on females. The patterns for male applicants match with those of overall income differences, while the significance of those for females is relatively low.

\subsubsection{Gender and Family Education}

\begin{figure}[H]
    \centering
    \begin{minipage}[H]{0.48\textwidth}
        \includegraphics[width=3in]{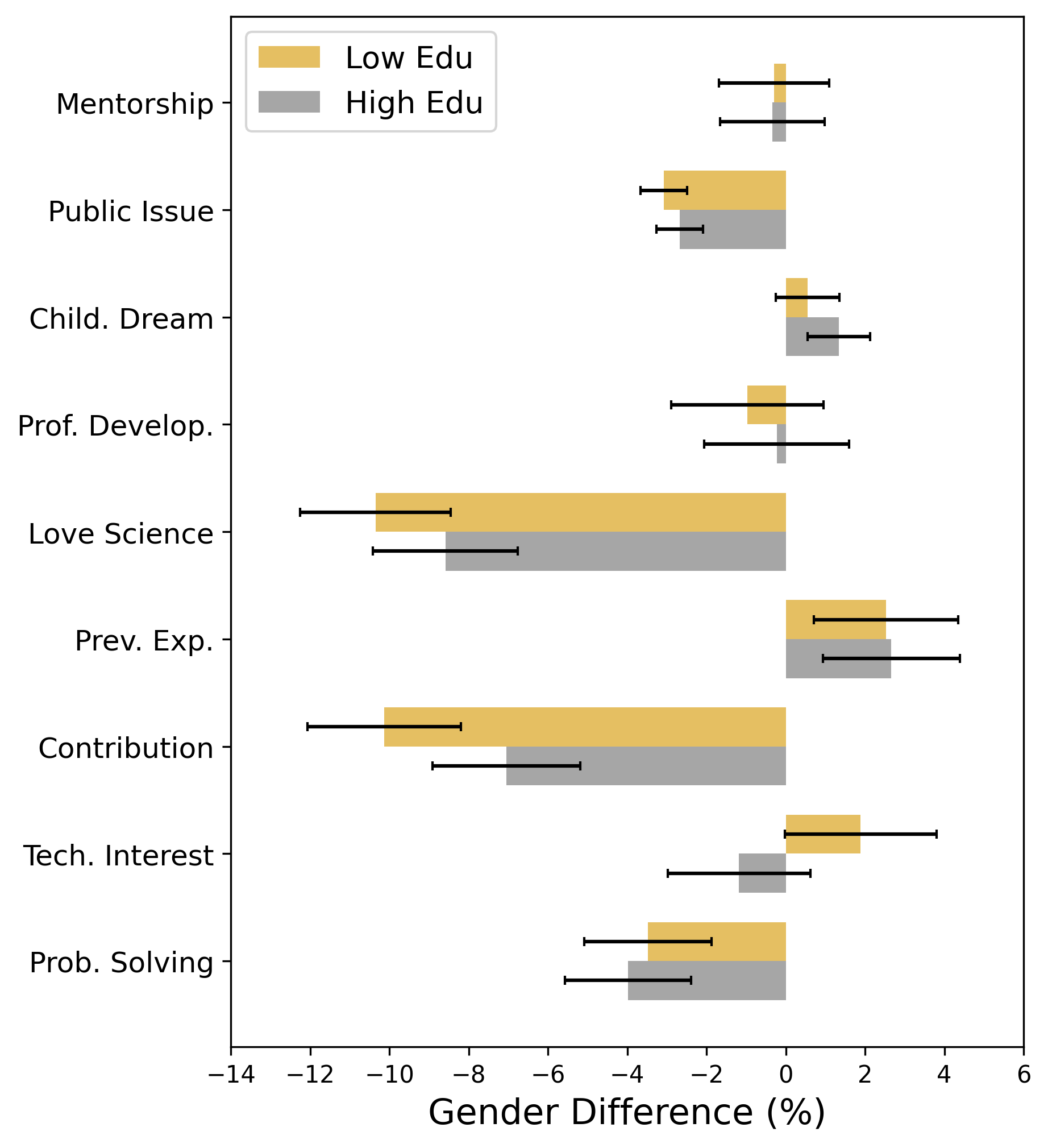} 
        \caption{Gender difference by education level. Error bars indicate 95\% confidence intervals.}
        \label{fig:gender_by_edu}
    \end{minipage}
    \begin{minipage}[H]{0.48\textwidth}
        \includegraphics[width=3in]{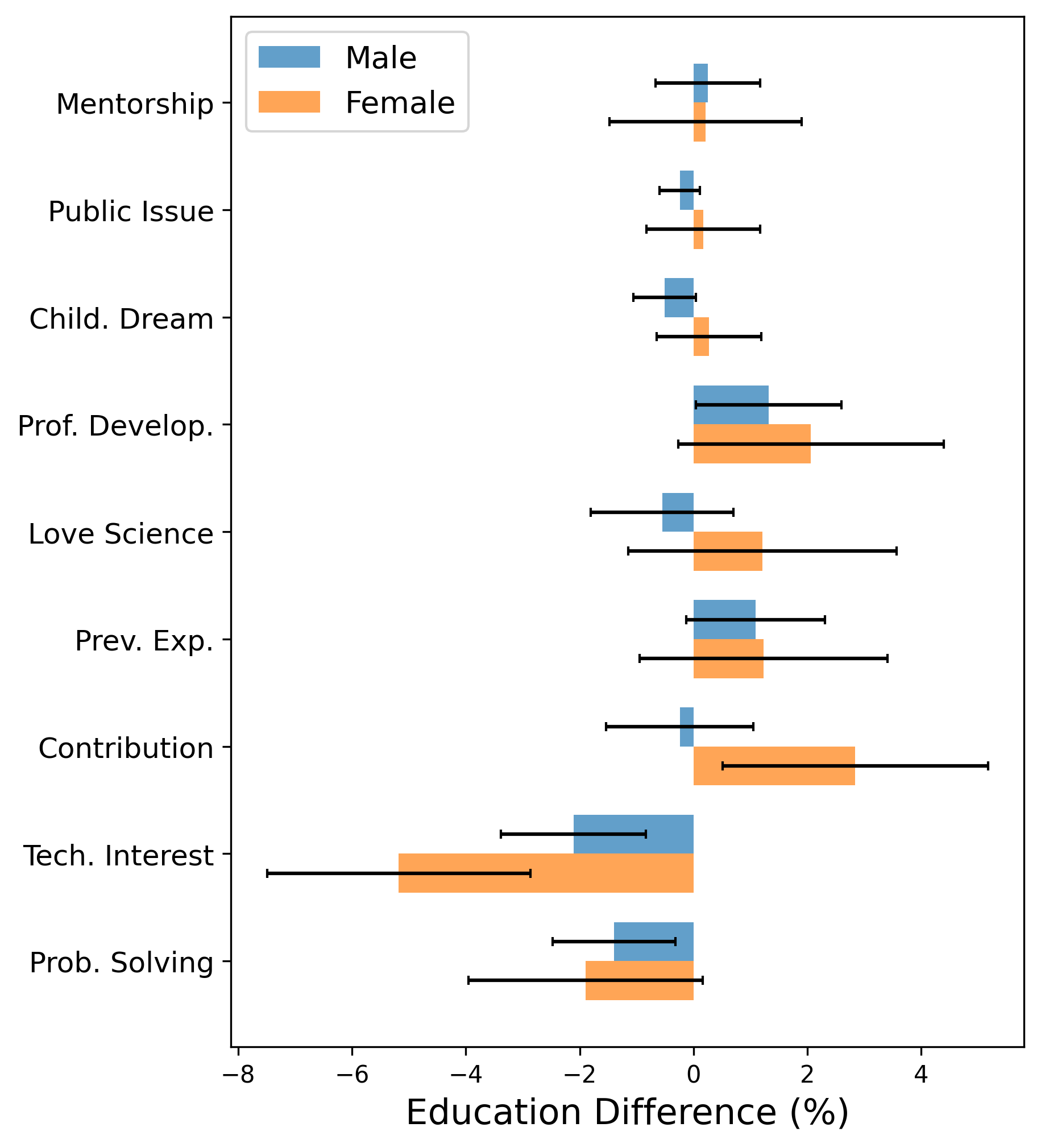} 
        \caption{Education difference by gender. Error bars indicate 95\% confidence intervals.}
        \label{fig:edu_by_gender}
    \end{minipage}
\end{figure}

We first compare the gender differences for applicants from different family-education subgroups. From Figure \ref{fig:gender_by_edu}, we observe that  gender-based differences remain consistent regardless of family education levels. As was also found in the earlier discussion of the interaction between gender and family income, the exception to the above lies only in the topic of ``childhood dream''. In this case as well, while there is a gender-based difference in the proportion of applicants that mention this topic, the difference is only significant in the high-education group. One possible explanation is that families with higher education levels or better economic status are generally able to provide their children with relevant STEM-related exposures and guidance at a young age, thus, help them establish aspirations in engineering. 

We next compare the education differences for female and male applicants, as summarized in Figure \ref{fig:edu_by_gender}. We make a number of observations: First, while for both genders, applicants with higher family educational level are more likely to mention the topic of ``technical interest'', the influence of this factor seems to be much larger in the female subgroup. Second, while no significant difference between family education levels in the proportional of applicants mentioning the topic of ``love of science'' and ``make contribution'' was observed for the entire applicant pool, Figure \ref{fig:edu_by_gender} shows that there is indeed a difference when looking at female applicants only - in their case, a higher family education level is related to a higher likelihood of mentioning both these topics.

\subsubsection{Family Income and Family Education}

\begin{figure}[!ht]
    \centering
        \includegraphics[width=3.5in]{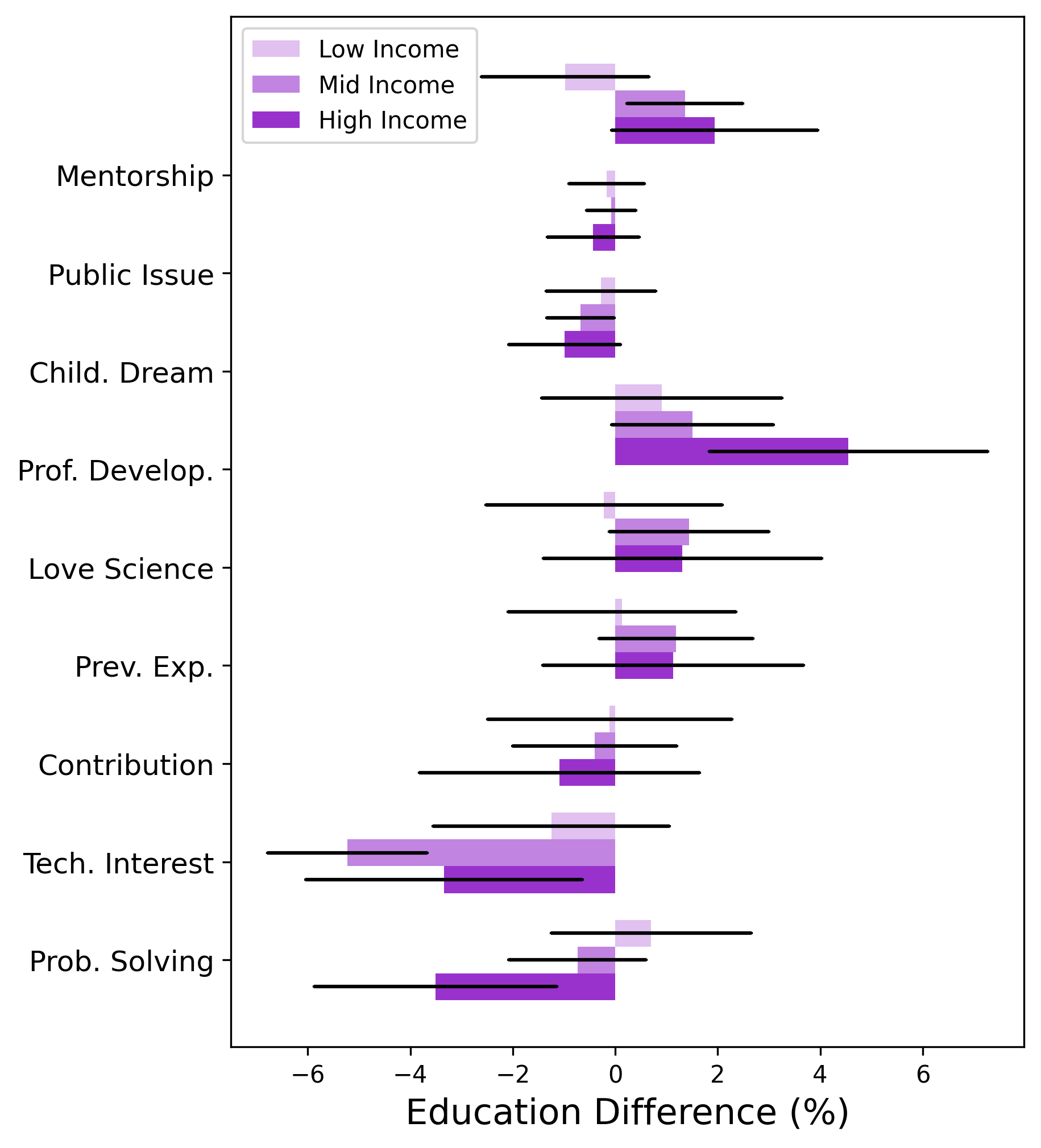} 
        \caption{Education difference by income level. Error bars indicate 95\% confidence intervals.}
        \label{fig:edu_by_income}
\end{figure}

Figure \ref{fig:edu_by_income} shows the impact of the family's education level on the frequency of the different motivational factors, for applicants at each of the three income level groups (low, mid, and high). When it comes to low-income applicants, family education backgrounds are shown to have no statistically significant impact on students' expressed reasons for studying engineering. In contrast, some education-based differences emerge in the mid- and high-income applicant groups. In particular, increased education level increases the frequency of the topics of ``technical interest" and ``problem solving", whereas lower education is associated with increased frequency of the topics of ``mentorship" and ``professional development". The results tell a story that family education background does change the way an applicant perceives and thinks about engineering as a field of study, but the impact is notable only if their family's income is average or higher. Otherwise, the influences attributed to the family's educational background might be dominated by those resulting from financial factors.

\begin{figure}[!ht]
    \centering
        \includegraphics[width=6in]{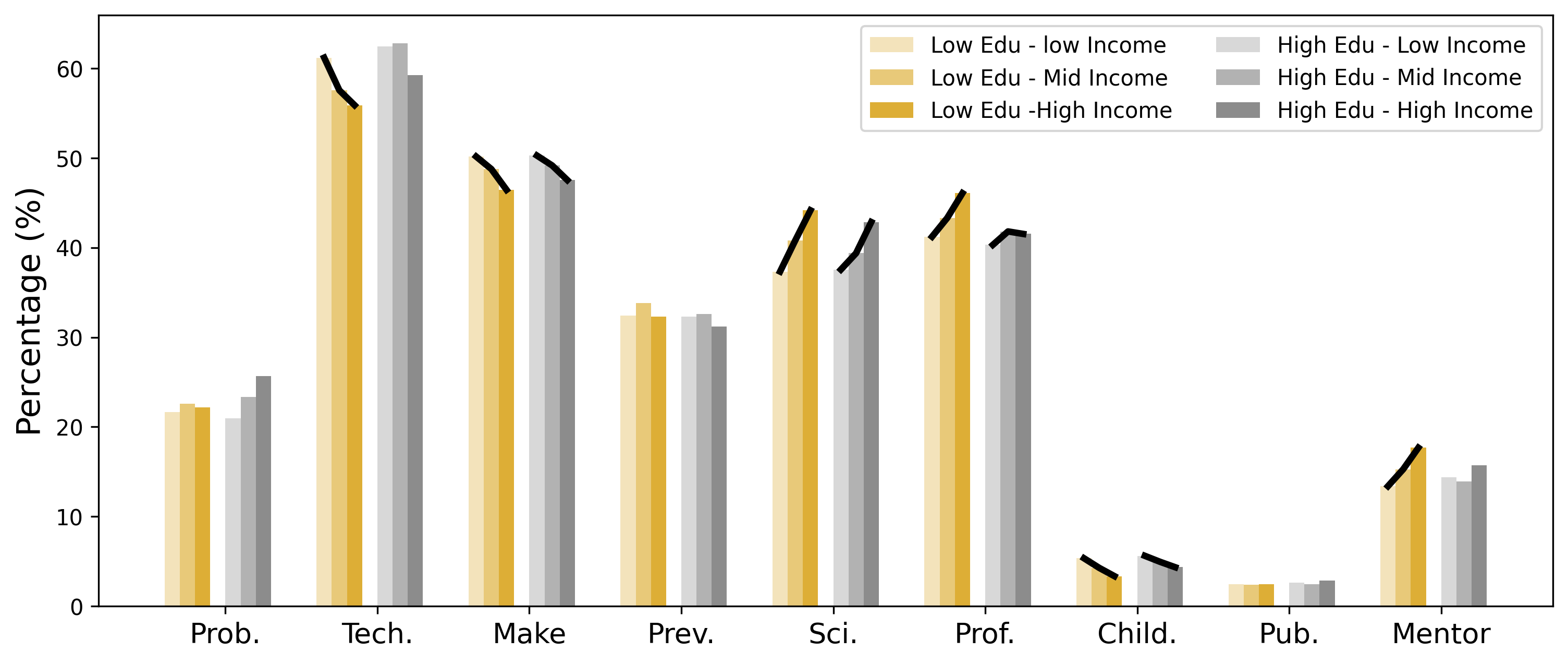} 
        \caption{Topic distribution by income and education. Education differences that are significant at a confidence level of 95\% are highlighted with trend lines.}
        \label{fig:income_by_edu}
\end{figure}

Figure \ref{fig:income_by_edu} presents the distribution of topics across different income levels, for each of the two educational level subgroups - low and high. In most cases, the significance of differences in motivational factors according to income are higher for low-education applicants than for applicants of better family education levels. As discussed, a family's economical advantage can be translated into educational resources for their offspring. However, our analysis shows that if parents have good education themselves, the impact of family socio-economic status can be at least partially offsetted because parents are able to provide guidance and act as role models for their children without any financial cost. 

\section{Conclusion} \label{sec:conclusion}
\subsection{Contribution and limitations}
Despite decades of efforts to attract more women to STEM programs, especially engineering and computer science, women remain underrepresented in engineering classrooms, their numbers dwindling further in engineering graduate studies and professional careers. Understanding students' motivations for applying to engineering programs can provide insight on if and how male and female applicants differ in their reasons for considering engineering as an educational and career path. 

Our presented approach uses applicants' text response combined with census data to extract motivational factors that drive students to pursue engineering as a field of study, which can benefit future talent recruitment in STEM. The novelty of our research approach is two-fold. First, we used an innovative hybrid of the LDA and the Word2Vec model for text clustering. The proposed model takes advantage of the LDA's interpretability and the Word2Vec's capability of precisely capturing semantic information, and is thus able to generate text clustering results of great reliability without extensive human intervention. Second, in addition to gender, we also considered applicants' nationality and their family income and education, which have been shown to have significant impact on the educational attainment of the next generation.  

Due to the lack of relevant data, we do not investigate the impact of family income and education on international students and do not consider other important factors such as ethnicity, which are shown by previous research to impact students' decisions regarding higher education \cite{ohland2008persistence}. In addition, we do not have access to final admission decisions, which could have served as a proxy for the educational attainment of the applicants themselves. Despite these shortcomings, our analysis leads to a number of meaningful comparisons and insights as well as recommendations for improving recruitment approaches for women in engineering.

\subsection{Summary of Findings}
We provide empirical evidence showing that there exist significant differences in motivational factors due to gender, nationality, and family income and education. 

First, female applicants are concerned more with the practical and social implications of engineering, while males establish aspirations about engineering through previous experiences and early childhood wishes. 

Second, a novel comparison of Canadian and international students found that Canadian students put greater emphasis on intrinsic factors, such as personal interest and professional development, and less focus on interpersonal relationships such as mentorship and social impact. We attribute this difference to cultural differences; over three quarters of the international students in our sample originated from Asia, the Middle East, and North-Africa, which are broadly known to be collectivist cultures, in contrast with the more individualistic Canadian culture \cite{triandis2018individualism, schwarts1994}. 

Third, family income and education backgrounds are shown to be significantly related to students' motivations for pursing engineering education. Students from families with high socio-economic status and advanced educational attainment generally have access to resources that allow them to receive information and guidance in related fields \cite{bowles2002inheritance, gould2011does}. Accordingly, in our findings, high-income applicants were more likely to refer to their love of problem solving and science, interest in professional development, and mentorship as sources of motivation. In contrast, applicants from low-income families were more likely to refer to their interests in technology and childhood dreams and  focus more on contributing to society with the help of their knowledge and skills. We found smaller differences due to family education level. 

Finally, when combined, these demographic factors influence students' perceptions and decisions about engineering in a complicated way. Among them, gender has the most robust impact as gender-based differences broadly persist regardless of applicants' nationality, family income and education levels. Family income and education levels are strongly correlated with each other and have similar impact on an applicant's motivation. In addition, the relationship between family income and motivations for pursuing engineering is stronger in males than females. 

\subsection{Implications}

These empirical findings have important implications for policy makers and educators. Our results suggest that in order to attract more students to engineering programs, efforts should be taken to expose young students to relevant activities to foster their interests in technology and science. In addition, engineering programs' marketing and recruitment activities should emphasize the usefulness of the knowledge and skills students would be able to gain through studying engineering, and highlight that as future engineers, they will be able to tackle real-world challenges, have good careers and play significant roles in addressing public issues.

In order to reduce the gender gap in STEM, strategies specifically tailored for females should be taken into consideration. From our findings we infer that, besides pointing at new technologies and the promising career paths, which are motivators equally attractive to males and females, engineering programs need to promote and highlight factors specifically targeted at females such as the practical significance and social impact of engineering. Females need to be aware of the significant contributions they would be able to make as professional engineers. Furthermore, incorporating more engineering examples into high school science education and establishing mentorship programs are also potential measures to enhance females' participation in STEM.

Our research has shown that the disadvantages due to lower family education and socio-economic backgrounds translate into measurable differences in applicants' interests in and perceptions of engineering. To mitigate this inequity, and to increase the number of under-represented minorities in engineering \cite{ASEE2019}, students from low income and low education families could be supported with services including but not limited to consulting, informal meetings, and mentorship to enable them to participate in extra-curricular activities in related fields.


\bibliographystyle{IEEEtran}
\bibliography{paper.bib}

\newpage

 




\vfill

\end{document}